\documentclass[prd,twocolumn,superscriptaddress,showpacs,nofootinbib,amsmath,amssymb]{revtex4}

\usepackage{bm}
\usepackage{amsfonts}
\usepackage{latexsym}
\usepackage[latin1]{inputenc}
\usepackage{graphicx}
\usepackage{amsmath}
\usepackage{rotating}
\usepackage{epsfig}

\newcommand{\mb}[1]{\mbox{\boldmath $#1$}}

\def \nn  {\nonumber}
\def \met {\mbox{g}}
\def \metb{\mbox{\bf g}}

\begin{document}

\def\jnl@style{\it}
\def\aaref@jnl#1{{\jnl@style#1}}

\def\aaref@jnl#1{{\jnl@style#1}}

\def\aj{\aaref@jnl{AJ}}                   
\def\apj{\aaref@jnl{ApJ}}                 
\def\apjl{\aaref@jnl{ApJ}}                
\def\apjs{\aaref@jnl{ApJS}}               
\def\apss{\aaref@jnl{Ap\&SS}}             
\def\aap{\aaref@jnl{A\&A}}                
\def\aapr{\aaref@jnl{A\&A~Rev.}}          
\def\aaps{\aaref@jnl{A\&AS}}              
\def\mnras{\aaref@jnl{MNRAS}}             
\def\prd{\aaref@jnl{Phys.~Rev.~D}}        
\def\prl{\aaref@jnl{Phys.~Rev.~Lett.}}    
\def\qjras{\aaref@jnl{QJRAS}}             
\def\skytel{\aaref@jnl{S\&T}}             
\def\ssr{\aaref@jnl{Space~Sci.~Rev.}}     
\def\zap{\aaref@jnl{ZAp}}                 
\def\nat{\aaref@jnl{Nature}}              
\def\aplett{\aaref@jnl{Astrophys.~Lett.}} 
\def\apspr{\aaref@jnl{Astrophys.~Space~Phys.~Res.}} 
\def\physrep{\aaref@jnl{Phys.~Rep.}}      
\def\physscr{\aaref@jnl{Phys.~Scr}}       

\let\astap=\aap
\let\apjlett=\apjl
\let\apjsupp=\apjs
\let\applopt=\ao

\title[Coupling of Radial and Axial non-Radial Oscillations of
Compact Stars] {Coupling of Radial and Axial non-Radial
Oscillations of Compact Stars: Gravitational Waves from first-order
Differential Rotation}

\author{Andrea Passamonti}
\affiliation{Institute of Cosmology and Gravitation, University
of Portsmouth, Mercantile House, Portsmouth PO1 2EG, UK}
\affiliation{Department of Physics, Aristotle University of
Thessaloniki 54124, Greece}
\author{Marco Bruni}
\affiliation{Institute of Cosmology and Gravitation, University
of Portsmouth, Mercantile House, Portsmouth PO1 2EG, UK}
\affiliation{Dipartimento di Fisica, Universit\`a  degli Studi di  Roma  ``Tor Vergata",
via della Ricerca Scientifica 1, 00133 Roma, Italy}
\author{Leonardo Gualtieri}
\affiliation{Centro Studi e Ricerche E. Fermi, Compendio Viminale, 00184 Rome, Italy and 
Dipartimento di Fisica ``G. Marconi'', Universit\`a di Roma
``La Sapienza''/Sezione INFN ROMA 1, P.le Aldo Moro 2, I-00185 Roma,
Italy}
\author{Alessandro Nagar}
\affiliation{Dipartimento di Fisica, Politecnico di Torino, Corso Duca degli
Abruzzi 24, 10129 Torino, Italy and INFN sez. di Torino, Via P. Giuria 1, Torino, Italy}
\author{Carlos F. Sopuerta}
\affiliation{Institute for Gravitational Physics and Geometry,
Center for Gravitational Wave Physics, and Department of Astronomy \&
Astrophysics, Penn State University, University Park, PA 16802, USA}
%

\date{\today}

\begin{abstract}
        We  investigate the  non-linear coupling between  radial   and   non-radial oscillations of
        static  spherically  symmetric neutron stars as a possible  mechanism for the generation of
        gravitational waves that may lead to observable signatures. In this paper we concentrate on
        the axial sector of the  non-radial perturbations.  By using a multi-parameter perturbative
        framework we introduce a complete description of the non-linear coupling between radial and
        axial non-radial  oscillations; we  study the gauge   invariant character of the associated
        perturbative   variables  and  develop  a   computational  scheme  to evolve the non-linear
        coupling perturbations in the time domain.  We present results of simulations corresponding
        to different  physical  situations  and  discuss the dynamical behaviour of this non-linear
        coupling.  Of particular interest is the occurrence of signal amplifications in the form of
        resonance phenomena when a frequency  associated  with the radial pulsations is close  to a
        frequency  associated  with  one of the  axial  $w$-modes of  the star. Finally, we mention
        possible extensions  of  this work  and improvements towards more astrophysically motivated
        scenarios.
\end{abstract}


\maketitle

\section{Introduction}
Gravitational wave astronomy is currently developing as a new way of
studying the cosmos, with a number of different experiments starting
operating or being developed: earth-based interferometers (LIGO~\cite{ligo},
VIRGO~\cite{virgo}, GEO600~\cite{geo} and TAMA~\cite{tama}), resonant bars
(EXPLORER, AURIGA, NAUTILUS, ALLEGRO and NIOBE; see~\cite{bars}) and spheres such as
MiniGRAIL~\cite{minigrail}, as well as the laser interferometer space
antenna LISA~\cite{lisa}.  The scientific success of these detectors
in producing new physical and astrophysical knowledge depends on the
amount of {\em a priori} available theoretical understanding about the
different sources of gravitational waves.  It is then of outmost importance to
have good theoretical descriptions of the systems that are likely to be
observed by the ongoing gravitational-wave experiments.

Due to the weak character of the gravitational interactions there is
only a few number of systems that can generate gravitational waves
detectable for the experiments just mentioned.  In general, the
dynamics of these systems involves strong gravitational fields that
need to be described, using different levels of approximations, by
general relativity.  This is the case in the modeling of compact
objects such as neutron stars and supernov\ae\ core collapse, where
non-linearity plays a role in the dynamics.  For these systems,
relativistic perturbation theory is an excellent approach.  The linear
theory has been used for a long time to study their oscillations and
instabilities~\cite{Andersson:2002ch,Kokkotas:2002ng}.  However, we
know relatively little about non-linear dynamical effects (mainly due
to numerical
studies~\cite{Sperhake:2001xi,2002PhRvD..65b4001S,2002ApJ...571..435M,2002PhRvD..65h4039L,2003ApJ...591.1129A,Font:2001ew,Stergioulas:2003ep,Dimmelmeier:2005zk}).
Despite the fact that strong non-linear effects require a fully
non-perturbative approach, it is reasonable to expect that there are
interesting physical phenomena that only involve a mild non-linearity
for which a second-order treatment should be perfectly suited.  In
this sense, a non-linear perturbative approach is therefore timely and
may lead to a better understanding of the mechanisms of generation of
gravitational waves.

An interesting scenario to study is a neutron star which is
oscillating radially and non-radially.  At first order radial
oscillations of a spherical star don't emit {\it per se} any
gravitational waves, but they can drive and possibly amplify
non-radial oscillations and then produce gravitational radiation to a
significant level. In addition one may expect the appearance of
non-linear harmonics, which may also come out at lower frequencies
than the linear modes~\cite{Zanotti:2004kp,Dimmelmeier:2005zk}, where
the Earth-based laser interferometers have a higher sensitivity.
An additional motivation is that there are a number of studies aiming
at investigating if non-radial oscillations of stars can be excited by
external sources (see e.g.\ \cite{Gualtieri:2001cm,Poisson:1993vp}).
However, our idea is to investigate whether non-radial oscillations
can be driven or even amplified through coupling by an internal radial
oscillation, regardless of the presence of an external source.  These
non-linear processes can occur, for instance, in a proto-neutron star
that is still pulsating.  A mainly radial pulsation could for example
drive the non-radial oscillations, either naturally present, or
excited through fall-back accretion.

In recent papers~\cite{Bruni:2002sm,Sopuerta:2003rg} a non-linear
multi-parameter perturbative formalism has been developed, which can
be applied to the study of the non-linear dynamics of the oscillations
of neutron stars, as shown in~\cite{Passamonti:2004je} where it has
been further developed to study a particular second-order effect: the
coupling of radial and non-radial first order perturbations of a
compact spherical star.  One of the main ideas behind this work is that
it is very convenient to treat the two sets of perturbations, radial
and non-radial, as separately parametrized and then to study, at the
next perturbative order, the way in which they couple, which
corresponds to a particular sector of the second-order perturbations,
the one formed by the product of radial and non-radial perturbations.
In~\cite{Passamonti:2004je}, the equations describing the coupling
were obtained for the case of polar non-radial perturbations and
gauge-invariant variables were also found (having fixed the gauge for
the radial ones for simplicity).

In this paper, we extend this study to the case of axial perturbations.
At first order, axial perturbations are decoupled from fluid perturbations,
but at the next perturbative order,
the terms describing the coupling are driven by the radial pulsations.
Here we discuss the results of simulations in the time domain that show
how the coupling mechanism works and under which conditions we may have
a situation in which this coupling produces amplifications due to resonant 
phenomena.

The plan of this paper is the following: In Section~\ref{framework} we
describe the perturbative framework we use in this work.  In
Section~\ref{ingredients} we introduce all the necessary ingredients
required by the perturbative scheme in order to study the coupling of radial
and axial non-radial oscillations of neutron stars.  In
Section~\ref{numerics} we describe the structure of the computational
framework we have implemented to solve the perturbative equations in the time domain.
In particular, we discuss in detail the construction of initial data and the
numerical treatment of the boundary conditions.
In Section~\ref{results} we present the results of our simulations for two
different physical scenarios: (i) the scattering of a gravitational wave
packet by a differentially rotating star, and (ii) the non-linear coupling
in a differentially rotating and radially oscillating compact star.
In particular we discuss the appearance of amplifications due to resonances
of the system.  We conclude in Section~\ref{conclusions},
where we summarize the main results of this paper and discuss potential
extensions of the approach we have used.  In the Appendices we give some proofs
of the gauge invariance character of some of the perturbative quantities
(Appendix~\ref{gaugeinvariance}), expressions for the source terms in
the perturbative equations describing the coupling
(Appendix~\ref{sourceterms}), and the equations for radial
perturbations in terms of the {\em tortoise} fluid coordinate
(Appendix~\ref{equationstortoise}).

Throughout this work we use the following conventions: We use Greek
letters to denote spacetime indices; capital Latin letters for indices
in the time-radial part of the metric; lower-case Latin indices for
the spherical sector of the metric.  We use physical units in which
$8\pi G = c = 1\,$.

\section{Perturbative Framework \label{framework}}

In order to investigate the effects of the coupling of linear axial
perturbations and first order radial oscillations we use a general
two-parameter framework recently developed
\cite{Bruni:2002sm,Sopuerta:2003rg}. In essence, this consists of
separately parametrizing the radial and non radial oscillations.  We
have presented the formalism for the coupling between radial and polar
non-radial perturbations in \cite{Passamonti:2004je}. In this paper we
develop and apply the formalism to study the coupling arising from
axial perturbations, following the same general lines.  In the
following we recall the general two-parameter framework. However, to
fix the ideas, we will refer explicitly to its application to the case
of stellar oscillations.

The physical spacetime of the oscillating star we consider is
represented as a non-linear perturbation of a static spherical
background stellar model, with metric $\mb{\metb}^{(0,0)}\,$.  The
expansion of the physical metric $\mb{\metb}$ around this background
has the following form:
\begin{equation}
\met^{}_{\alpha\beta} = \met^{(0,0)}_{\alpha\beta} +
\lambda\,\met^{(1,0)}_{\alpha\beta} +
\epsilon\,\met^{(0,1)}_{\alpha\beta} +
\lambda\epsilon\,\met^{(1,1)}_{\alpha\beta} +
O(\lambda^2,\epsilon^2)\,, \label{initialmetric}
\end{equation}
where $\lambda$ and $\epsilon$ are the expansion parameters associated
with radial and non-radial perturbations respectively.  We use
superscripts $(I,J)$ to denote perturbations of order
$\lambda^I\epsilon^J$.  Therefore, the terms $\mb{\metb}^{(1,0)}$ and
$\mb{\metb}^{(0,1)}$ in~(\ref{initialmetric}) represent, respectively,
first-order radial and non-radial perturbations, while
$\mb{\metb}^{(1,1)}$ is the non-linear contribution due to their
coupling. This is the quantity we are interested in this paper, while
we will neglect the self-coupling terms of order $\lambda^2$ and
$\epsilon^2$.

Any physical quantity can be expanded as the metric
in~(\ref{initialmetric}).  In particular, the energy-momentum tensor
has the expansion:
\begin{equation}
\label{tmunu}
T^{}_{\alpha\beta} = T^{(0,0)}_{\alpha\beta} + \lambda T^{(1,0)}_{\alpha\beta}+
\epsilon T^{(0,1)}_{\alpha\beta} + \lambda\epsilon T^{(1,1)}_{\alpha\beta}
+O(\lambda^2,\epsilon^2) \,.
\end{equation}
Let us now analyze the structure of the field equations 
\begin{equation}
\mb{E}\left[\,\metb\,,\mb{\psi}^{}_A\,\right] = \mb{G}\left[\,\metb\,\right] -
\mb{T}\left[\,\metb\,,\mb{\psi}^{}_A\,\right] = 0 \label{efes}
\end{equation}
arising from the perturbative expansion (see e.g.~\cite{Wald:1984cw}).
Here $\mb{G}$ denotes the Einstein tensor, $\mb{T}$ the
energy-momentum tensor and $\mb{\psi}^{}_A$ (A=$1,\dots$) the matter
variables.  Introducing (\ref{initialmetric}) and  
(\ref{tmunu}) into Eq.~(\ref{efes}) we obtain the
perturbative equations at each order: 
\begin{widetext}
\begin{align}
& \mb{E} \left[\,\metb\,,\mb{\psi}^{}_A\,\right] =
\mb{E}^{(0,0)}\left[\,\mb{\metb}^{(0,0)}\,,\mb{\psi}_A^{(0,0)}\,\right]
+ \lambda \mb{E}^{(1,0)}
\left[\,\mb{\metb}^{(1,0)}\,,\mb{\psi}_A^{(1,0)}\,\right] +
\epsilon \mb{E}^{(0,1)}
\left[\,\mb{\metb}^{(0,1)}\,,\mb{\psi}_A^{(0,1)}\,\right] \nn
\\
& \quad +\lambda\epsilon \mb{E}^{(1,1)}\left[\,\mb{\metb}^{(1,1)}\,,
\mb{\psi}_A^{(1,1)}\;\left|\;
\mb{\metb}^{(1,0)}\otimes\mb{\metb}^{(0,1)}\,,
\mb{\psi}_A^{(1,0)}\otimes\mb{\psi}_A^{(0,1)}\,,
\mb{\metb}^{(1,0)}\otimes\mb{\psi}_A^{(0,1)}\,,
\mb{\psi}_A^{(1,0)}\otimes\mb{\metb}^{(0,1)} \,\right] +
O(\lambda^2,\epsilon^2)=0 \,. \right. \label{total}
\end{align}
\end{widetext}
This equation has to be satisfied for arbitrary values of the
expansion parameters, therefore each perturbative term must vanish
independently.  This naturally gives rise to an iterative scheme,
where at each non-linear order one has to solve inhomogeneous
equations with source terms containing the solutions of the lower
order equations.

The lowest order in (\ref{total}), $\mb{E}^{(0,0)} = 0$, gives the
equations for the background, in our case the
Tolman-Oppenheimer-Volkoff (TOV) equations.  The other terms
$\mb{E}^{(I,J)} = 0$ represent the perturbative equations of order
$\lambda^I\epsilon^J\,$.  The structure of the operators
$\mb{E}^{(I,J)}$, with $(I,J)\neq(0,0)$, is such that they act linearly
on every term specified between the square brackets, while they are 
non-linear functions of the background quantities $\metb^{(0,0)}$ and
$\mb{\psi}_A^{(0,0)}$.  At first order in $\lambda$, we obtain the
equations describing the radial perturbations of the TOV background,
$\mb{E}^{(1,0)}=0$, whereas axial non-radial perturbations come from
the first-order terms in $\epsilon$:
\begin{equation}
\mb{E}^{(0,1)}\left[\,\mb{\metb}^{(0,1)}\,,\mb{\psi}_A^{(0,1)}\,\right]=0\,.
\label{Enonr}
\end{equation}
Finally, the equations of  order $\lambda\epsilon$ describe the
radial non-radial coupling:
\begin{align}
& \mb{E}^{(1,1)}\biggl[\,\mb{\metb}^{(1,1)}\,,\mb{\psi}_A^{(1,1)}~\left|~
\mb{\metb}^{(1,0)}\otimes\mb{\metb}^{(0,1)}\,,
\mb{\psi}_A^{(1,0)}\otimes\mb{\psi}_A^{(0,1)}\,, \right. \nn
\\
& ~~~~~~~~~~ \mb{\metb}^{(1,0)}\otimes\mb{\psi}_A^{(0,1)}\,,
\mb{\psi}_A^{(1,0)}\otimes\mb{\metb}^{(0,1)} \,\biggr] = 0 \,, \label{EinCoup}
\end{align}
where the vertical bar separates the unknowns of order
$\lambda\epsilon$ from the quantities computed in the previous
iteration of the perturbative scheme, which form the source terms of
the equations.

In order to solve these equations, it is important to take into
account that the operator $\mb{E}^{(1,1)}$ acts on
$(\mb{\metb}^{(1,1)}\,,\mb{\psi}_A^{(1,1)})$ in Eq.~(\ref{EinCoup}) in
the same way as $\mb{E}^{(0,1)}$ acts on
$(\mb{\metb}^{(0,1)}\,,\mb{\psi}_A^{(0,1)})$ in Eq.~(\ref{Enonr}).
The reason is that both operators come from the linearization of the
Einstein tensor operator acting on axial non-radial perturbations.
Then, using the linearity of $\mb{E}^{(1,1)}$, we can define
\begin{equation}
\mb{L}^{}_{\rm NR}\left[\,\cdot\,\right] \equiv \mb{E}^{(1,1)}\left[\,\cdot\, |\,
\mb{0}\,\right]  = \mb{E}^{(0,1)}\left[\,\cdot\,\right]  \,,
\end{equation}
as the axial non-radial perturbation operator. Hence, equation (\ref{EinCoup})
can be written in the form:
\begin{align}
& \mb{L}^{}_{\rm NR}\left[\,\mb{\metb}^{(1,1)}\,,\mb{\psi}_A^{(1,1)}\,\right]=
\mb{S}\biggl[ \mb{\metb}^{(1,0)}\otimes\mb{\metb}^{(0,1)}\,, \nn
\\
& \mb{\psi}_A^{(1,0)}\otimes\mb{\psi}_A^{(0,1)}\,,
\mb{\metb}^{(1,0)}\otimes\mb{\psi}_A^{(0,1)}\,,
\mb{\psi}_A^{(1,0)}\otimes\mb{\metb}^{(0,1)} \,\biggr]\,.\label{Einsep}
\end{align}
This particular structure of the equations is very important in
solving them by using time-domain numerical methods.  Indeed, we can
easily extend a numerical code that solves the equations for
first-order axial non-radial perturbations simply by adding the source
terms on the right-hand side of (\ref{Einsep}).  Later on we will
discuss additional advantages of this structure.

\section{Perturbative scheme for stellar oscillations\label{ingredients}}
In this section we introduce the basic ingredients needed to study the
coupling of radial and axial perturbations of relativistic stars: the
background description (Sec.~\ref{background}), the first-order radial
and axial perturbations (Secs.~\ref{radialpert} and~\ref{axialpert}),
and finally the equations describing the coupling of the first-order
perturbations (Sec.~\ref{couplingpert}).  To simplify our expressions
we denote the background quantities with an overbar,
e.g. $\bar{\met}^{}_{\alpha\beta}= \met^{(0,0)}_{\alpha\beta}$ and
$\bar{T}^{}_{\alpha\beta}= T^{(0,0)}_{\alpha\beta}\,$.

\subsection{The static spherically-symmetric background\label{background}}
The background is an equilibrium configuration described by a static
spherically-symmetric metric (see, e.g.,~\cite{Misner:1973cw}), a TOV
model.  The line-element is:
\begin{equation}
\bar{\met}^{}_{\alpha\beta}dx^\alpha dx^\beta = -e^{2\Phi}dt^2
+ e^{2\Lambda}dr^2+r^2(d\theta^2+\sin^2\theta d\phi^2) \,,
\end{equation}
where $\Phi$ and $\Lambda$ are functions of $r$ only.  This is a
solution of Einstein's equations with a perfect-fluid energy-momentum
tensor
\begin{equation}
\bar{T}^{}_{\alpha\beta} = (\bar{\rho}+\bar{p})\, \bar{u}^{}_\alpha
\bar{u}^{}_\beta + \bar{p}\, \bar{\met}^{}_{\alpha\beta}\,,
\end{equation}
where the four-velocity of the fluid is
$\bar{u}^{}_\alpha=\left(-e^\Phi,0,0,0\right)$, and in general $\rho$ and
$p$ denote the energy density and the pressure.  The mass function
$m(r)$ is introduced through the metric function $\Lambda$:
$e^{-2\Lambda(r)}=1-2m(r)/r\,$. The TOV equations of stellar
equilibrium read:
\begin{eqnarray}
\Phi^{}_{,r} & = & \frac{m+4\pi r^3\bar{p}}{r(r-2m)} = -\frac{\bar{p}^{}_{,r}}
{\bar{\rho}+\bar{p}} \,, \label{Phieq}
\\
m^{}_{,r}& = & 4\pi r^2 \bar{\rho} \,. \label{meq}
\end{eqnarray}
Specifying the Equation of State (EoS) of the stellar equilibrium
configuration yields a 1-parameter family of solutions of the
equations (\ref{Phieq},\ref{meq}), which is typically parametrized by
the value of the central density $\bar{\rho}^{}_c = \bar\rho(r\!=\!0)\,$.
Here we consider a configuration described by an adiabatic,
barotropic EoS generalization of the Newtonian polytropic
EoS:
\begin{equation}
\bar{p} = K \bar\rho^{\Gamma} \,, \label{eos}
\end{equation}
where $K$ is a constant and $\Gamma$ the polytropic index.  In
particular, we take $K\!=\!100$ km$^2$, $\Gamma \!=\! 2\,$,
$\bar\rho^{}_c = 3\times 10^{15}$ g/cm$^{3}$. These values yield a
stellar configuration with mass $M=1.26M^{}_{\odot}$ and radius 
$R=8.86$ km.  We denote by $\bar{c}^{}_s=(d{\bar{p}}/d{\bar{\rho}})^{1/2}$ the
speed of sound.

\subsection{Radial Perturbations of the TOV Model\label{radialpert}}
Radial perturbations (first investigated by
S. Chandrasekhar~\cite{Chandrasekhar:1964pr,Chandrasekhar:1964tc}; see
also~\cite{Kokkotas:2000up}), by definition, preserve the spherical
symmetry of the TOV background.  They can be completely described in
terms of the Lagrangian displacement of a fluid element around its
equilibrium position.  Alternatively, in the {\em radial} gauge, they
can be written in terms of two metric functions:
\begin{eqnarray}
g^{(1,0)}_{\alpha\beta}dx^\alpha dx^\beta & = &
e^{2\Phi}(rS^{(1,0)} - 2\eta^{(1,0)})dt^2 \nonumber
\\
& + & re^{2\Lambda}S^{(1,0)} dr^2  \,.
\end{eqnarray}
The radial perturbations of the fluid variables in this gauge are:
\begin{eqnarray}
u_\alpha^{(1,0)}dx^\alpha & = & e^\Phi (\textstyle{\frac{r}{2}}S^{(1,0)} -
\eta^{(1,0)}) dt + e^{\Lambda}\gamma^{(1,0)}dr \ ,\\
\delta \rho^{(1,0)} &=& \bar\rho\,\omega^{(1,0)}\ , \\
\delta p^{(1,0)}    &=& \bar{c}_s^2 \bar\rho\,\omega^{(1,0)} \,.
\end{eqnarray}
To simplify the equations 
we use the following perturbative quantities: {\em metric variables}:
the metric perturbations $\eta^{(1,0)}$ and $S^{(1,0)}\,$; {\em matter
variables}: the fluid velocity perturbation $\gamma^{(1,0)}$ and the
enthalpy perturbation, which is related to the density perturbations
through the relation:
\begin{equation}
H^{(1,0)} = \frac{\bar{c}_s^2 \bar\rho}{\bar\rho+\bar p}
\omega^{(1,0)}  \,.
\end{equation}
The dynamics of radial pulsations is then governed by the following system
of PDEs:
\begin{widetext}
\begin{eqnarray}
 H_{,t}^{(1,0)} & = & -\bar{c}_s^2 e^{\Phi-\Lambda} \left\{
\gamma^{(1,0)}_{,r} + e^{2\Lambda}\left[\left(1-\frac{1}
{\bar{c}_s^2}\right) \left(4\pi r\bar p + \frac{m}{r^2} \right)
+ \frac{2}{r} e^{-2\Lambda} - 4\pi(\bar\rho + \bar p) \right]
\gamma^{(1,0)}\right\}\,, \label{eq:H10_ev}
\\
\gamma_{,t}^{(1,0)} & = & - e^{\Phi+\Lambda}
\left\{ e^{-2\Lambda} H^{(1,0)}_{,r} + 4\pi r (\bar\rho + \bar p) H^{(1,0)} +
\frac{1}{2}\left(1+8\pi r^2 \bar p\right)S^{(1,0)}\right\}\,, \label{eq:gam_t}
\\
S_{,t}^{(1,0)} & = & -8\pi(\bar\rho +\bar p) e^{\Phi+\Lambda}
\gamma^{(1,0)}\,, \label{chi_t}
\end{eqnarray}
\end{widetext}
which was derived in~\cite{Passamonti:2004je} using the formalism
of~\cite{Gerlach:1979rw,Gerlach:1980tx,Martin-Garcia:1998sk,Gundlach:1999bt,Martin-Garcia:2000ze}
(see Section~\ref{axialpert} below).  A similar system was obtained
in~\cite{Ruoff:2000nj} using the well-known ADM 3+1
formalism~\cite{ArnDesMis:62,York:79}. Having solved the above system,
the metric variable $\eta^{(1,0)}$ can then be obtained from the
following relation
\begin{equation}
\eta^{(1,0)}_{,r} = 4\pi r(\bar\rho+\bar p)
\left[r S^{(1,0)}+\left(1+ \frac{1}{\bar{c}^2_s}\right)H^{(1,0)}
\right] e^{2 \Lambda} \,. \label{eq:eta_cn}
\end{equation}
Finally, our variables have to satisfy the Hamiltonian constraint, which takes
the following form
\begin{equation}
S^{(1,0)}_{,r} = e^{2\Lambda}\biggl[ \left(8\pi r\bar\rho - \frac{2}{r} +
\frac{2 m}{r^2} \right) S^{(1,0)}+8 \pi \frac{\bar\rho + \bar p}{\bar{c}_s^2}
H^{(1,0)}\biggr] \,. \label{eq:S10_cn}
\end{equation}
Boundary conditions must be fixed at the center and at the stellar
surface $r\!=\!R$. The vanishing at the surface of the first-order
Lagrangian pressure perturbation, $\Delta p^{(1,0)}(R)=0$, 
leads to the following condition on $\gamma^{(1,0)}$
\cite{Passamonti:2004je}:
\begin{equation}
(\bar\rho + \bar p)\,\bar{c}^2_s ( r^2
e^{-\Lambda} \gamma^{(1,0)})^{}_{,r} |^{}_{r=R}= 0\,.  \label{BC10_srf}
\end{equation}
The behaviour of $S^{(1,0)}$ and $H^{(1,0)}$ on the surface can be
derived from the general evolution equations (\ref{eq:H10_ev}) and
(\ref{chi_t}). In the \emph{radial} gauge there is still a residual
gauge degree of freedom, which can be used for setting to zero all the
pure gauge perturbations of the vacuum external
spacetime~\cite{Martin-Garcia:2000ze}. For stellar models where the
density vanishes on the stellar surface, this gauge choice induces
also $S^{(1,0)}$ and $\eta^{(1,0)}$ to vanish on that
surface~\cite{Martin-Garcia:2000ze}.  At $r=0$, regularity conditions
imply
\begin{eqnarray}
S^{(1,0)} & = & r S^{(1,0)}_o(t)+ O(r^3) \,, \label{bcradial1}
\\
\eta^{(1,0)} & = & \eta^{(1,0)}_o(t)+O(r^2) \,,\label{bcradial2}
\\
H^{(1,0)} & = & H^{(1,0)}_o(t)+O(r^2) \,,\label{bcradial3}
\\
\gamma^{(1,0)} & = & r \gamma^{(1,0)}_o(t)+ O(r^3)\,. \label{bcradial4}
\end{eqnarray}

\subsection{Axial Perturbations of the TOV Model\label{axialpert}}
Axial oscillations of a static star were first investigated by Thorne
and
Campolattaro~\cite{Thorne:1967th,Thorne:1968tc,Campolattaro:1970ct},
and have been the subject of many subsequent studies (see,
e.g.~\cite{Chandrasekhar:1991cf,Andersson:1996ak}).

Non-radial perturbations are nicely described in the formalism
introduced by Gerlach and
Sengupta~\cite{Gerlach:1979rw,Gerlach:1980tx} and further developed by
Gundlach and
Mart\'{\i}n-Garc\'{\i}a~\cite{Martin-Garcia:1998sk,Gundlach:1999bt,Martin-Garcia:2000ze}
(which we denote as GSGM formalism). This was originally introduced to
study first order perturbations of a general time-dependent
spherically symmetric stellar background.  In particular we can apply
it to our case, where the background is the static TOV metric.

\subsubsection{Summary of the GSGM formalism for axial perturbations}
The key point in the GSGM formalism is the fact that the background
manifold is a warped product $M^2\times S^2$, where $S^2$ denotes the
2-sphere and $M^2$ is a two-dimensional Lorentzian manifold.  The
metric can be written as the semidirect product of a general
Lorentzian metric on $M^2$, $\bar{g}^{}_{AB}$, and the unit curvature
metric $\gamma^{}_{ab}$ on $S^2$:
\begin{equation}
\bar{\met}^{}_{\alpha\beta} =
\left(\begin{array}{cc} \bar{g}^{}_{AB} & 0 \\
0 & r^2\gamma^{}_{ab} \end{array} \right) \,. \label{met22}
\end{equation}
Hereafter $x^A$ denote the coordinates
on $M^2$, $x^a$ the coordinates on $S^2$ and $r=r(x^A)$ is a function on
$M^2$ that coincides with the invariantly defined radial (area) coordinate of
spherically-symmetric spacetime. A vertical bar is used to denote
the covariant derivative on $M^2$ and a semicolon to denote that
on $S^2$. The metrics are covariantly conserved, i.e. $
\bar{g}^{}_{AB|C}=\gamma^{}_{ab:c}=0$ .  One can introduce the completely
antisymmetric covariant unit tensors on $M^2$ and on $S^2$,
$\bar{\epsilon}^{}_{AB}$ and $\epsilon^{}_{ab}$ respectively.

The energy-momentum tensor has the same block diagonal structure
as the metric:
\begin{equation}
\bar{T}^{}_{\alpha\beta}= \textrm{diag}
\left(\bar{\rho}\bar{u}^{}_A\bar{u}^{}_B+\bar{p}\bar{n}^{}_A\bar{n}^{}_B\,,\,
r^2 \bar{p}\, \gamma^{}_{ab} \right)\,, \label{tblock}
\end{equation}
where $\bar{u}^{}_A$ are the non-zero components of the fluid velocity,
$\bar{u}^{}_\alpha=(\bar{u}^{}_A,0)\,$, and $\bar{n}^{}_A$ is a unit spacelike
vector given by
\begin{equation}
\bar{n}^{}_A = -\bar\epsilon^{}_{AB}\bar{u}^B~~\Rightarrow~~
\bar{n}^{}_A\bar{u}^A=0\,.
\end{equation}

Here we will focus on the axial non-radial perturbations
(see~\cite{Passamonti:2004je} for the polar ones).
The metric and energy-momentum axial perturbations can
be expanded in terms of the vector and tensor axial (or odd-parity)
spherical harmonics
\begin{equation}
S^{lm}_a = \epsilon_a{}^bY^{lm}_{:b}\,,~~~
S^{lm}_{ab} = S^{lm}_{a:b} + S^{lm}_{b:a} \,,
\end{equation}
where $Y^{lm}$ are the scalar spherical harmonics, in the following
form
\begin{eqnarray}
g^{(0,1)}_{\alpha\beta} &=&
\left(\begin{array}{cc} 0 & h_A^{lm} \, S^{lm}_a \\[2mm]
h_A^{lm} \, S^{lm}_a  &  h^{lm} S^{lm}_{ab}
\end{array}\right)\label{talbe1}\\
T^{(0,1)}_{\alpha\beta} &=&
\left(\begin{array}{cc} 0 & \delta t_A^{lm} \, S^{lm}_a \\[2mm]
\delta t_A^{lm} \, S^{lm}_a  & \delta t^{lm} S^{lm}_{ab}
\end{array}\right), \label{talbe2}
\end{eqnarray}
where $h_A^{lm}\,$ $h^{lm}\,$, $\delta t_A^{lm}\,$, and
$\delta t^{lm}$ are functions of $t$ and $r$ only. To simplify the
notation, in the following we drop the harmonic indices $(l,m)$.
As shown in~\cite{Gerlach:1979rw,Gerlach:1980tx}, a complete
set of axial gauge-invariant quantities is obtained by
taking the following combinations of $h^{}_A\,$, $h\,$,
$\delta t^{}_A\,$, and $\delta t\,$:
\begin{eqnarray}
k^{}_A & = & h^{}_A - h^{}_{\mid A } + 2 h \bar{v}^{}_A \,, \label{GIkA}
\\
L^{}_A & = & \delta t^{}_A - \bar{p}\, h^{}_A \,, \label{GILA}
\\
L   & = & \delta t - \bar{p}\, h \,, \label{GIL}
\end{eqnarray}
where $k^{}_A$ and $L^{}_A$ are defined for $l\ge 1\,,$ $L$ is defined for
$l\ge 2$, and $\bar{v}^{}_A= r^{-1}r^{}_{|A}\,$.  The only axial fluid
perturbation function arises from the expansion of the fluid velocity
\begin{equation}
u^{(0,1)}_{\alpha}=\left( 0 \,, \beta S^{}_a\right)\,, \label{deltau}
\end{equation}
where $\beta$ is a gauge invariant variable.  As it has been shown
in~\cite{Gundlach:1999bt}, for a perfect fluid the energy-momentum
gauge-invariant quantities take the form
\begin{equation}
L^{}_A = (\bar\rho + \bar{p})\beta \bar{u}^{}_A\,, \qquad L = 0 \,.
\end{equation}
Then, the equations for the perturbations can be decoupled in terms
of the following quantities~\cite{Gerlach:1979rw,Gerlach:1980tx}
\begin{equation}
\Psi = r^3 \, \bar{\epsilon}^{AB} \left( r^{-2} k^{}_A \right)_{\mid B} \,,
~~
\hat\beta = (\bar\rho+\bar{p})\beta
\label{Psibetadef}
\end{equation}
and their expressions are
\begin{align}
& \left[\frac{(r\Psi)^{}_{\mid A}}{r^2}\right]^{|A} -
\frac{(l-1)(l+2)}{r^3}\Psi = -16\pi\bar\epsilon^{AB}
(\hat\beta \bar{u}^{}_A)^{}_{\mid B}\,, \label{maseq}
\\
& u^A{\hat\beta}^{}_{\mid A} + (\bar\mu+2\bar{U})\hat\beta  =  0 \,,
\label{traseq}
\end{align}
where the background quantities $\bar{\mu}$ and $\bar{U}$ are
\begin{equation}
\bar{\mu} = \bar{u}^A{}^{}_{\mid A}\,,~~
\bar{U} = \bar{u}^A\bar{v}^{}_A = r^{-1}\bar{u}^Ar^{}_{\mid A}\,.
\end{equation}
We can solve these equations by prescribing initial data for $\hat\beta$ and
$\Psi$ at a time $t^{}_o$, that is, giving
$(\Psi^{}_o,\dot{\Psi}^{}_o,\hat\beta^{}_o)\,$.
Equation (\ref{traseq}) is an equation for $\hat\beta$ only that can
be solved independently from (\ref{maseq}). Its solution is constant
along the integral curves of $\bar{u}^A$ (see~\cite{Gundlach:1999bt}).
With the solution of (\ref{traseq}) we can solve the equation (\ref{maseq})
for $\Psi\,$, where $\hat\beta$ enters only in the source term.
Once we have solved equations (\ref{maseq},\ref{traseq}) for $\Psi$ and
$\hat\beta$, we can get the gauge-invariant metric perturbations $k^{}_A$
by means of the following relation:
\begin{equation}
(l-1)(l+2) k^{}_A = 16 \pi r^2 \hat\beta\, \bar{u}^{}_A -
\bar\epsilon^{}_{AB} (r\Psi)^{\mid B}\,. \label{mtr_form}
\end{equation}
We emphasize that the formalism as described so far applies to a
general time dependent, spherically symmetric background spacetime.

\subsubsection{GSGM formalism on the TOV background\label{axialperttov}}
In the case of our TOV static background, we have:
\begin{equation}
\bar{u}^A = (e^{-\Phi},0)\,,~~
\bar{n}^A = (0,e^{-\Lambda})\,,~~
\bar\mu = \bar{U} = 0\,.  \label{unst}
\end{equation}
Then, the equations for the gauge-invariant axial perturbations $\Psi$
and $\hat\beta$ are
\begin{widetext}
\begin{align}
& -\Psi^{}_{,tt} + \Psi^{}_{,r_\ast r_\ast} +
e^{2\Phi}\left[ 4\pi(\bar{p}-\bar\rho) + \frac{6M}{r^3} -
\frac{l(l+1)}{r^2} \right]\Psi +
16\pi r e^{2\Phi+\Lambda}\left[ e^{-2\Lambda}\hat\beta^{}_{,r} +
\left(4\pi r\bar p + \frac{M}{r^2}\right)
\hat\beta\right] = 0 \,, \label{Psi01maseq}
\\
& \hat\beta^{}_{,t} = 0 \,, \label{traseq01}
\end{align}
\end{widetext}
where, to simplify the equations, we have introduced the {\em tortoise}
coordinate
\begin{equation}
dr^{}_\ast = e^{\Phi-\Lambda} dr\,.
\end{equation}
Equation (\ref{traseq01}) tells us that $\hat\beta$ is not a dynamical
variable and hence, at first order, is typically of no interest and
can be consistently set to
zero~\cite{Thorne:1967th,Chandrasekhar:1991cf,Andersson:1996ak}.
In this sense, it is said that first-order axial perturbations
do not depend on the fluid motion. 

As we discuss later in Section~\ref{sec:ID}, $\beta^{}_{lm}$ can however
be taken to represent the harmonic component $lm$ of a first order
differential rotation profile, and as such plays a very interesting
role in the coupling, at order $\lambda\epsilon$, with radial
oscillations.

Then, if we consider a non-vanishing $\hat\beta\,$, the general
solution of~(\ref{Psi01maseq}) is the combination of (i) the solutions
of the homogeneous associate equation that describes the only
dynamical degrees of freedom of axial perturbations, i.e. the
gravitational waves, and (ii) a particular (static) solution that is
related to the dragging of inertial frames induced by a non-vanishing
$\hat\beta$, as discussed in~\cite{Thorne:1967th}.

Outside the star, equation~(\ref{Psi01maseq}) reduces
to the well-known Regge-Wheeler equation
\begin{equation}
-\Psi^{}_{,tt} + \Psi^{}_{,r_\ast r_\ast} -
V^{}_l(r) \Psi = 0 \,. \label{RW01eq}
\end{equation}
where
\begin{equation}
V^{}_l(r) = \left(1-\frac{2M}{r}\right)\left(\frac{l(l+1)}{r^2}
-\frac{6M}{r^3}\right) \,, \label{rwpotential}
\end{equation}
and $M$ is the gravitational mass of the TOV star.
In~(\ref{RW01eq}), $r^{}_\ast$ is the usual Regge-Wheeler tortoise
coordinate $r^{}_\ast\equiv r+2M\ln[r/(2M)-1]\,$.
To complete the description of the first-order axial perturbations
we need to discuss the boundary conditions at the origin,
at the stellar surface, and at infinity.
The requirement of regularity of the perturbations at the origin yields
\begin{eqnarray}
\hat\beta \sim r^{l+1}\,,  \qquad
\Psi \sim r^{l+1}  \,. \label{BC01_orig}
\end{eqnarray}
The matching conditions on the stellar surface imply the
continuity of metric variable $\Psi\,$, of its time derivative,
and of the following quantity~\cite{Martin-Garcia:2000ze}:
\begin{equation}
e^{-\Lambda}\left(r^{-3}\Psi\right)^{}_{,r} - 16\pi r^{-2}\,
\hat\beta \,. \label{SC01}
\end{equation}
In the case of a barotropic equation of state, the pressure and
mass-energy density vanish on the surface, and the condition
(\ref{SC01}) reduces to the continuity of $\Psi^{}_{,r}$.
At infinity we can simply use Sommerfeld outgoing wave conditions.

The master function $\Psi$ is related to the emitted
power in gravitational waves at infinity (see, e.g.~\cite{Nagar:2005ea}):
\begin{equation}
\frac{dE}{dt} = \sum_{l,m}P_{l m}=\frac{1}{16\pi} \sum^{}_{l,m}
\frac{l(l+1)}{(l-1)(l+2)}|\dot{{\Psi}}_{lm}|^2\, , \label{gwpower}
\end{equation}
where we have explicitly restored the harmonic indices $(l,m)$ and
the overdot denotes time differentiation with respect to the
Schwarzschild coordinate time.

\subsection{Non-linear Coupling between radial and non-radial perturbations\label{couplingpert}}
To derive the equations that describe the non-linear coupling between
first-order radial and axial non-radial oscillations we are going to
follow the procedure used in the polar case~\cite{Passamonti:2004je}.
In brief, the idea is to obtain the equations for the coupling from
the equations for the axial perturbations (\ref{maseq}) and
(\ref{traseq}), by using the following formal perturbative expansion:
\begin{equation}
\met^{}_{\alpha\beta} = \met^{(0)}_{\alpha\beta} + \epsilon \,
\met^{(1)}_{\alpha\beta}\,,\label{reorg}
\end{equation}
where $\met^{(0)}_{\alpha\beta}$ is a time dependent background composed of
our static TOV background with the radial oscillations, that is
$\met^{(0)}_{\alpha\beta} = \bar{\met}^{}_{\alpha\beta}
+\lambda\met_{\alpha\beta}^{(1,0)}\,$, and $\met^{(1)}_{\alpha\beta}$
represents axial non-radial perturbations containing the first-order
radial perturbations and the non-linear coupling terms, that is
$\met^{(1)}_{\alpha\beta} = \met^{(0,1)}_{\alpha\beta}+\lambda
\met^{(1,1)}_{\alpha\beta}\,$.  As we have already mentioned in
Sec.~\ref{framework} this is due to the fact that the first-order axial and
the coupling perturbations have essentially the same structure and
are governed by the same type of equations.

The metric and energy-momentum axial perturbation of order $(1,1)$ can
be expanded in odd-parity spherical harmonics as in the first-order
case [see equations~(\ref{talbe1}), (\ref{talbe2})].  To distinguish
them we use superscripts $(I,J)$ denoting the perturbative order
$\lambda^I\epsilon^J\,$.  In this way, the $(0,1)$ metric and
energy-momentum perturbations are denoted by $h_A^{(0,1)}\,$,
$h^{(0,1)}\,$, $t_A^{(0,1)}\,$, and $t^{(0,1)}\,$; and the $(1,1)$
ones by $h_A^{(1,1)}\,$, $h^{(1,1)}\,$, $t_A^{(1,1)}\,$, and
$t^{(1,1)}\,$.  Likewise, the velocity perturbations of order $(1,1)$
are written as $u^{(1,1)}_{\alpha}= (0\,,\beta^{(1,1)}S^{}_a)\,$.

Then, the gauge-invariant quantities that we construct by using the ansatz
of (\ref{reorg}) are:
\begin{eqnarray}
k_A^{(1,1)} & = & h_A^{(1,1)} - h^{(1,1)}_{\mid A } + 2h^{(1,1)}\bar{v}^{}_A
\label{ka_11_exp}  \,,
\\
L_A^{(1,1)} & = & \delta t_A^{(1,1)} - \bar{p} h_A^{(1,1)}-
\bar{\rho}\bar{c}^2_s\omega^{(1,0)} h_A^{(0,1)} \,, \label{LA_11_exp}
\\
L^{(1,1)} & = & \delta t^{(1,1)} - \bar{p} h^{(1,1)} -
\bar{\rho}\bar{c}^2_s\omega^{(1,0)} h^{(0,1)} \,.  \label{L_11_exp}
\end{eqnarray}
Comparing expressions (\ref{ka_11_exp})-(\ref{L_11_exp}) with
expressions (\ref{GIkA})-(\ref{GIL}) we notice that they differ
because of the presence of the products of first-order perturbations
in the $(1,1)$ perturbations in~(\ref{LA_11_exp},\ref{L_11_exp}) (see
also~\cite{Passamonti:2004je}).  The gauge invariance of the $(1,1)$
quantities (\ref{ka_11_exp})-(\ref{L_11_exp}) is shown in
Appendix~\ref{gaugeinvariance}.

We can also construct the coupling perturbations $\hat\beta^{(1,1)}$
and  $\Psi^{(1,1)}$ in the same way as $\hat\beta^{(0,1)}=\hat\beta$ and
$\Psi^{(0,1)}=\Psi$ in (\ref{Psibetadef}) (see Appendix~\ref{gaugeinvariance}
for more details).   We find that they satisfy the following equations:
\begin{widetext}
\begin{align}
& -\Psi^{(1,1)}_{,tt} +  \Psi^{(1,1)}_{,r_\ast r_\ast}
+ e^{2\Phi}\left[4\pi(\bar{p}-\bar\rho) + \frac{6 M}{r^3}
- \frac{l(l+1)}{r^2}\right]\Psi^{(1,1)}
+ 16\pi r e^{2\Phi+\Lambda}\left[ e^{-2\Lambda}\hat\beta_{,r}^{(1,1)} +
\left(4\pi r\bar p + \frac{M}{r^2}\right)
\hat\beta^{(1,1)}\right] = e^{2 \Phi} \Sigma^{}_{\Psi}\,, \label{Psi11maseq}
\\
& \hat \beta^{(1,1)}_{,t} = e^{\Phi}\Sigma^{}_{\beta}\,, \label{traseq11}
\end{align}
\end{widetext}
where $\Sigma^{}_{\Psi}$ and $\Sigma^{}_{\beta}$ are source terms made out of
products of radial and axial perturbations.  Their expressions can be
found in Appendix~\ref{sourceterms}.  In the stellar exterior we do
not have fluid perturbations and the radial perturbations vanish due
to Birkhoff's theorem.  Hence, the equations for the exterior do not
have source terms and then (\ref{Psi11maseq}) reduces to the
Regge-Wheeler equation for $\Psi^{(1,1)}$ [same as for $\Psi^{(0,1)}$,
see Eqs.~(\ref{RW01eq},\ref{rwpotential})].  Therefore, the emitted
power in gravitational waves at infinity due to the (1,1)
perturbations is given by equation~(\ref{gwpower}), substituting
$\dot{\Psi}^{}_{lm}$ by $\dot{\Psi}^{(1,1)}_{lm}\,$.  Regarding the
boundary conditions for $\hat\beta^{(1,1)}$ and $\Psi^{(1,1)}\,$, at
the origin they are the same as the ones for $\hat\beta^{(0,1)}$ and
$\Psi^{(0,1)}$ [see Eq.~(\ref{BC01_orig})]; at infinity we use
outgoing wave Sommerfeld conditions (see below); at the stellar surface,
both $\Psi^{(1,1)}$ and its time derivatives are continuous.  The
matching conditions~\cite{Martin-Garcia:2000ze} also imply the
continuity of the following quantity
\begin{align}
& e^{-\Lambda}\left[\Psi^{(1,1)}_{,r} + \left(\frac{3}{2}\Psi^{(0,1)}
-\frac{r}{2}\Psi^{(0,1)}_{,r}\right)S^{(1,0)}\right] \nn
\\
& + e^{-\Phi}\gamma^{(1,0)}\Psi^{(0,1)}_{,t} + 16\pi r\hat\beta^{(1,1)}\,.
\label{Sfbc11}
\end{align}
This condition, which must be satisfied by the initial data, is
preserved by the evolution equations~\cite{Martin-Garcia:2000ze}.

\section{Setup For Numerical Computations\label{numerics}}
In this section we describe the setup we have used in the numerical simulations
of the coupling of radial and axial non-radial oscillations of relativistic
stars.  A more detailed description of the numerical implementation can be found
in~\cite{Passamonti:2005ph}.

\subsection{Structure of the Numerical Codes}

It is clear from the characteristics of the perturbative framework
described in Sec.~\ref{framework} that the we have to implement the
following computational steps. (i) Construction, once and for all, of
the TOV background; then, at each timestep: (ii) independent
integration of the first-order radial and non-radial perturbative
equations; (iii) updating, using the results of (ii), of the source
terms in the perturbative equations describing the coupling; (iv) use
the source computed in (iii) to integrate the equations for the
coupling variables.  In what follows we describe how these different
steps are carried out.

To solve the TOV equations (\ref{Phieq},\ref{meq}) of hydrostatic
equilibrium, we prescribe the value of the central density
$\bar\rho^{}_c$ and integrate from the center to the surface, where the
pressure vanishes.  The integration is carried out by using a fourth-order
Runge-Kutta method.

Radial perturbations obey equations
(\ref{eq:H10_ev})-(\ref{eq:S10_cn}).  It is well known (see,
e.g.~\cite{Sperhake:2001si}) that an accurate numerical integration
requires some attention at the stellar surface. Indeed, the low values
that the speed of sound takes near the surface reduce the convergence
rate of a second-order numerical scheme to first order.  This problem
can be solved by introducing, near the stellar surface, a {\em tortoise fluid}
coordinate $x$~\cite{Ruoff:2001ux}:
\begin{equation}
dr = \bar{c}^{}_s \, dx ~~~\Longrightarrow ~~~
\partial^{}_r = (1/\bar{c}^{}_s)\partial^{}_x \,. \label{tort_fluid}
\end{equation}
A uniform grid based on $x$ ($x$\emph{-grid}) will provide the
necessary resolution near the surface by simulating a non uniform
mesh in the $r$ coordinate. The transformation~(\ref{tort_fluid})
has to be implemented in the TOV equations and in the equations for
the radial perturbations (see Appendix~\ref{equationstortoise}).

The are two ways to deal with the equations for the radial perturbations.
One is to use a free evolution scheme based on a purely hyperbolic formulation,
which means to solve the equations (\ref{H10_ev_x})-(\ref{S_t_x}) and to
monitor errors through the Hamiltonian constraint~(\ref{S10_cn_x}).
The second way is to use an elliptic-hyperbolic scheme, where we integrate
Eqs.~(\ref{H10_ev_x},\ref{gam_t_x}) for $H^{(1,0)}$ and $\gamma^{(1,0)}\,$,
and solve the Hamiltonian constraint~(\ref{S10_cn_x}) for $S^{(1,0)}\,$.
In this second scheme the Hamiltonian constraint is satisfied at every
time-step by construction.  In both schemes, we need to integrate
Eqs.~(\ref{H10_ev_x},\ref{gam_t_x}), for which we use a two-step
Mac-Cormack algorithm with a predictor-corrector step to provide
second-order accuracy both in space and time.   When the metric
perturbation $S^{(1,0)}$ is found from the evolution
equation~(\ref{S_t_x}) we use the same algorithm, whereas when it
is found from the Hamiltonian constraint~(\ref{S10_cn_x}) we use
a standard LU decomposition to solve the resulting tridiagonal
linear system.  Finally, the metric perturbation $\eta^{(1,0)}$
is found, at every time step, by integrating equation~(\ref{eta_cn_x})
by using the shooting method in order to implement the boundary
condition $\eta^{(1,0)}=0$ at the surface.

We use the second scheme for our computations, i.e. we integrate
numerically equations~(\ref{H10_ev_x})-(\ref{S10_cn_x}).  The
amplitude of the Hamiltonian constraint~(\ref{S10_cn_x}) remains
bounded by very small values for long-term evolutions and scales with
the grid resolution as expected in a second-order convergent numerical
evolution.  The degree of convergence in the $L^2$ norm for the
variables describing the radial perturbations 
are given in Table~\ref{tab_rad}.

\begin{table}
\caption{Convergence test for radial perturbations:
         $\sigma^{(0,1)}$ denotes the convergence rate in the
         $L^2$ norm. \label{tab_rad}}
\begin{ruledtabular}
\begin{tabular}{c|c c c c}
Variables:~~ & $~\gamma^{(0,1)}~$ & $~H^{(0,1)}~$  & $~S^{(0,1)}~$  & $~\eta^{(0,1)}~$   \\
\hline
$\sigma^{(0,1)}$   & 2.04   & 2.06   & 2.07   & 1.53
\end{tabular}
\end{ruledtabular}
\end{table}

\begin{figure}[t]
\begin{center}
\includegraphics[width=85mm]{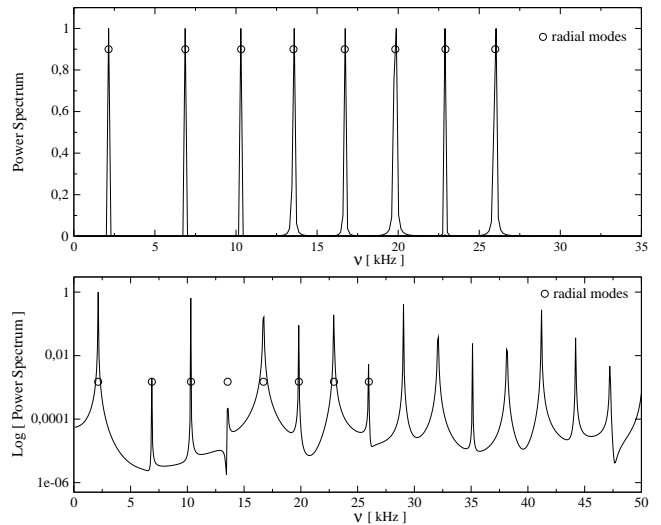}
\caption{Power spectrum of the radial fluid-velocity perturbation
$\gamma^{(1,0)}$. The \emph{upper panel} displays simultaneously
the FFT of eight different time evolutions.  In each evolution a single
radial mode has been excited.   In the \emph{lower panel}, radial
oscillations are excited with an initial Gaussian pulse.
The \emph{circles} denote the frequencies of the radial modes
determined from the eigenvalue problem~(see Section~\ref{sec:ID}).  Note
the different horizon scales used in the panels.
\label{radial_spectrum}}
\end{center}
\end{figure}

Another test of the numerical code for the radial perturbations
consists in comparing the mode frequencies available in the
literature, determined within a frequency domain approach, with those
we obtain by applying a Fast Fourier Transformation (FFT) to the
solution of our perturbative equations.
In Fig.~\ref{radial_spectrum}, we show the spectrum of the fluid
velocity perturbation $\gamma^{(1,0)}$, where the radial oscillations
have been excited with a selected eigenfrequency (\emph{upper panel})
and with an initial Gaussian pulse (\emph{lower panel}).  The radial
frequencies, excited during the time domain simulations, show an
agreement to better than $\sim 0.2 \,\%$ with the eigenfrequencies
determined in Section~\ref{sec:ID} with a frequency domain code, and with
the three known eigenfrequencies~\cite{Kokkotas:2000up}.

In the case of the first-order axial perturbations equations, as well
as in the case of the non-linear coupling, near the stellar surface we
don't have the problems exhibited by radial perturbations.  Therefore,
there is no need to introduce the {\em tortoise coordinate} and we can
use an evenly spaced grid in $r$ ($r$\emph{-grid}) for the $(0,1)$ and
$(1,1)$ perturbations.

In order to build the source terms for the coupling between
first-order radial and non-radial perturbations, we need to
interpolate the values of radial pulsations from the $x$\emph{-grid}
to the $r$\emph{-grid}.  We find a good agreement between the evolved
and interpolated radial variables when we use a $x$\emph{-grid} whose
resolution is twice that of the $r$\emph{-grid}.

We update in time the first-order axial perturbations by using 
equation~(\ref{Psi01maseq}) for the axial master variable
$\Psi^{(0,1)}$ and equation~(\ref{traseq01}) for the axial velocity
perturbation $\beta^{(0,1)}\,$.  The integration of~(\ref{traseq01})
is trivial: it yields a static profile for $\beta^{(0,1)}\,$.
For a nonzero velocity
perturbation, the master equation~(\ref{Psi01maseq}) contains a static
source term.  Then, the solution of~(\ref{Psi01maseq}) can be written
as: $\Psi^{(0,1)} = \Psi^{(0,1)}_h + \Psi^{(0,1)}_p\,$, where
$\Psi^{(0,1)}_h\,$ is the solution of the homogeneous equation and
describes the dynamical degrees of freedom of the gravitational field,
namely, it describes the gravitational wave content of the
non-radial oscillations.  $\Psi^{(0,1)}_p\,$ is a
particular solution of the full equations which can be chosen to be
time independent, and is related to the dragging of the
inertial frames due to the stellar rotation.  The homogeneous part
can be obtained numerically by using a standard Leapfrog scheme (see,
e.g.~\cite{1992nrfa.book.....P}).  The particular solution $\Psi^{(0,1)}_p$
satisfies the following ordinary differential equation:
\begin{widetext}
\begin{eqnarray}
\Psi^{(0,1)}_{p,rr}    + e^{2\Lambda}\left[4\pi r(\bar{p}-\bar\rho)
  + \frac{2M}{r^2}\right]\Psi^{(0,1)}_{p,r} & +  & e^{2\Lambda}\left[4\pi(\bar{p}-\bar\rho)
+ \frac{6M}{r^3} - \frac{l(l+1)}{r^2}\right] \Psi^{(0,1)}_{p}
\nn \\ {} &
+ & 16 \pi r e^\Lambda \left[\hat\beta_{,r}^{(0,1)} + e^{2\Lambda}\left(4\pi r \bar p
+ \frac{M}{r^2}\right)\hat\beta^{(0,1)}\right] = 0 \,, \label{Psi01prt}
\end{eqnarray}
\end{widetext}
which can be discretized in space using a second-order Finite Differences
approximation and written as a tridiagonal linear system.  Its
solution can then be found by using a standard LU
decomposition~\cite{1992nrfa.book.....P}.  The particular solution
$\Psi^{(0,1)}_p$ is related to the gauge-invariant metric component
$k_0^{(0,1)}$ through the relation (\ref{mtr_form}).  As it was argued
in~\cite{Thorne:1967th}, it describes the frame dragging of the
inertial frames due to the presence of the axial velocity
perturbation:
\begin{equation}
- \omega(r,\theta)r^2\sin^2\theta = \met_{t\phi}^{(0,1)} =
\sum^{}_{l,m} k_0^{(0,1)lm} S_{\phi}^{lm}\,,   \label{frmdr-rel}
\end{equation}
where the equality $k_0^{(0,1)lm} = h_0^{(0,1)lm}$ holds in the Regge-Wheeler
gauge or when the field is time independent. The metric perturbation
$k_0^{(0,1)lm}\,$, as well as the frame dragging harmonic component $\omega^{lm}\,$,
can also be determined directly from the following differential equation:
\begin{widetext}
\begin{equation}
e^{-2\Lambda}k_{0,rr}^{(0,1)} - 4\pi r(\bar\rho + \bar{p})
k_{0,r}^{(0,1)} + \left[8\pi(\bar\rho + \bar{p}) + \frac{4M}{r^3} -
\frac{l(l+1)}{r^2}\right] k_{0}^{(0,1)} = 16\pi e^{\Phi}\hat{\beta}^{(0,1)}
\,, \label{k0-fradr}
\end{equation}
\end{widetext}
which follows from~(\ref{Psibetadef}),~(\ref{mtr_form}). It can be
solved with the same method used for equation~(\ref{Psi01prt}).

The linear non-radial perturbations are the combination of a
gravitational wave degree of freedom and of a static solution related
to the dragging of the inertial frame.  They correspond, respectively,
to the dynamical solution $\Psi^{(0,1)}_{h}$ and to the particular
solution $\Psi^{(0,1)}_{p}$.  Our tests show that $\Psi^{(0,1)}_{h}$
and $k^{}_{0}$ have the expected second order convergence, while
$\Psi^{(0,1)}_{p}$ manifests a convergence of first order.  This lower
convergence rate is due to the discontinuity of $\Psi^{(0,1)}_{p,rr}$
at the stellar surface.

\begin{figure}[t]
\begin{center}
\includegraphics[width=85mm]{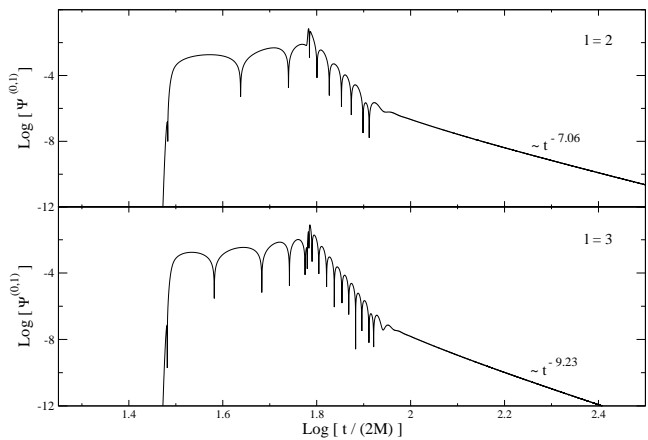}
\caption{\label{tail_axial}Gravitational-wave signal determined by the scattering 
of a Gaussian pulse on a TOV star. The exponentially damped sinusoidal oscillations 
corresponds to the $l=2$ ({\it top panel}) and to the $l=3$ ({\it bottom panel}) 
$w$-mode ringdown. The slope of the tail is in good agreement with the theoretical 
prediction. }
\end{center}
\end{figure}

\begin{figure}[t]
\begin{center}
\includegraphics[width=85mm]{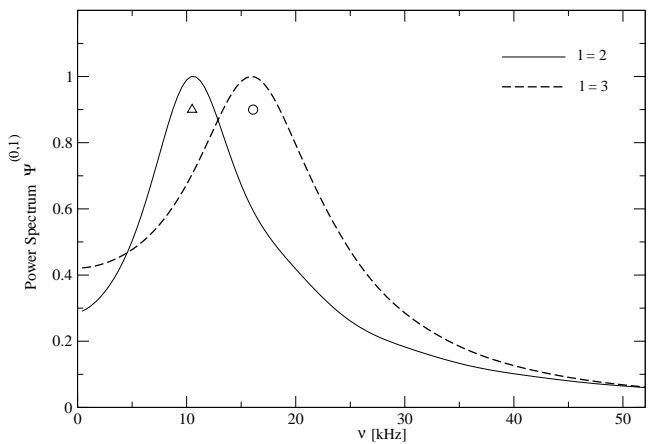}
\caption{Spectrum of the first order axial perturbations due to the scattering of
a Gaussian pulse on a TOV star. The peak correspond to the first $w$-mode excitation,
at frequencies $\nu^{}_w=10.5$ kHz ($l=2$) and $\nu^{}_{w} = 16.092$ kHz ($l=3$).
\label{spectrum_axial}}
\end{center}
\end{figure}

We have tested the performance of our numerical code for linear axial
perturbations by comparing the results of our simulations of the
scattering of an incident  gravitational wave onto a compact star with
results in the literature (see, e.g.~\cite{Ferrari:2000fk,Kojima:2005bw}).
For the quadrupolar case (see Fig.~\ref{tail_axial}), the scattered
gravitational signal exhibits the characteristic excitation of the
first $w$-mode, followed by
the ringing phase which is strongly damped by the emission of
gravitational radiation.
The behaviour of the signal at late times is known to be dominated by
the gravitational-wave tails due to back-scattering of the waves by
the spacetime curvature.  The theoretically expected time behaviour of
the tails is $\Psi^{(0,1)} \sim t^{-(2l+3)}$ (see~\cite{1995PhRvD..52.2118C}).
The behaviour we obtain from our simulations is shown in Fig.~\ref{tail_axial},
where by using a linear regression we obtain $\Psi^{(0,1)}_{h} \sim t^{-7.1}$
for $l=2$ and $ \Psi^{(0,1)}_{h} \sim t^{-9.26}$ for $l=3$.  We have checked
that these exponents approach the theoretical values as we increase the
evolution time.   The spectral properties of the solution are extracted by
Fourier transforming the signal.  In Fig.~\ref{spectrum_axial},
the power spectrum curves show the excitation of the lowest order \emph{w-modes}.
The frequencies associated with the peaks of these curves are in
excellent agreement with the frequencies computed by the frequency-domain code
used in~\cite{benhar-2004-70}.  Their values are:
$\nu^{}_{w} = 10.501$ kHz for $l=2$ and $\nu^{}_{w} = 16.092$ kHz for $l=3$.
The broad shape of the peaks is due to the short values of the \emph{w-modes} 
damping time, $\tau = 29.5$ $\mu$s for $l=2$ and $\tau = 25.2$ $\mu$s for 
$l=3$.

\begin{figure}[t]
\begin{center}
\includegraphics[width=85mm, height=65mm]{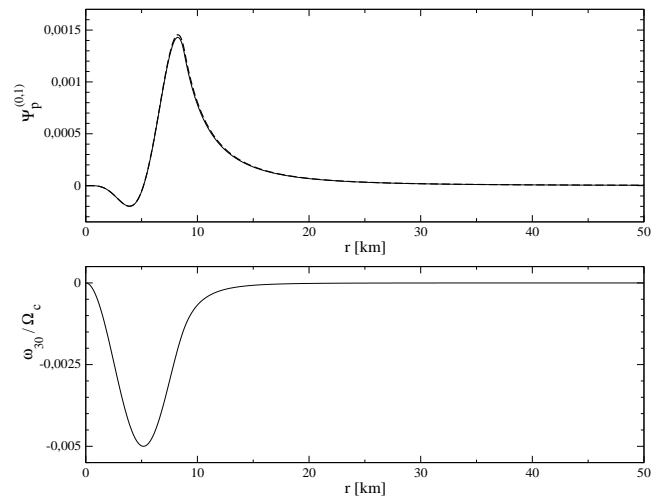}
\caption{ \label{frame_dragging}The stationary axial master function $\Psi^{(0,1)}_p$ relative to a nearly 
Newtonian $j$-rotation law with $A=15$ km and a period $T=10$ ms at the rotation axis
(top panel). The solution of the equation~(\ref{Psi01prt}) is showed with the
solid line while the dashed line denotes the solution found indirectly by first solving 
Eq.~(\ref{k0-fradr}) for the variable~$k_0^{(0,1)}$ and then using the definition~(\ref{mtr_form_stat}). 
The frame dragging function $\omega^{}_{30}$ determined by the initial value~(\ref{bet01_IC})
is displayed in the bottom panel.}
\end{center}
\end{figure}

In order to test the numerical solutions of the equation~(\ref{Psi01prt}) for
the axial master function $\Psi_p^{(0,1)}$ (for a particular profile of the axial 
velocity perturbation described in subsection~\ref{sec:ID}), we have compared the 
results with a different method.
This alternative method consists in solving first the equation~(\ref{k0-fradr}) 
for the metric perturbation $k^{}_0$ and then to obtain $\Psi_p^{(0,1)}$ from the
definition~(\ref{mtr_form}), which for the stationary case becomes
\begin{equation}
\Psi^{(0,1)}_p = \left(2 k_{0}^{(0,1)} - r k_{0,r}^{(0,1)}\right)
e^{-\left(\Phi + \Lambda \right)} \,. \label{mtr_form_stat}
\end{equation}
We have found that these two different computations agree to better than $2.3\,\%$
(see Figure~\ref{frame_dragging}).

The evolution equations for the non-linear coupling perturbations,
equations~(\ref{Psi11maseq},\ref{traseq11}), have the same structure
as the equations for linear axial perturbations, with the only
difference that the former contain source terms.  The source terms
$\Sigma^{}_{\Psi}$ and $\Sigma^{}_{\beta}$ [their expressions are given in
equations~(\ref{Spsi11}) and~(\ref{Sbe11}) respectively] are
discretized by second-order centered finite difference approximations
in the internal grid points and by second-order one-sided Finite
Difference approximations at the origin and at the stellar surface.
The computational domain for the axial master equation~(\ref{Psi11maseq}) 
is the entire spacetime, but the sources~$\Sigma^{}_{\Psi}$ and 
$\Sigma^{}_{\beta}$ have support only in the {\em interior} spacetime 
(inside the star).  We have tested different numerical schemes to solve 
the equation for the master function~$\Psi^{(1,1)}$.  We have found that 
the scheme that provides the best accuracy depends on the physical setting 
that we are considering.  In the case we are studying the coupling between 
the radial pulsations and differential rotation, our calculations are more
accurate when we use an upwind evolution algorithm for the interior
and a leapfrog scheme for the exterior.  In the case in which we study
the scattering of an axial gravitational wave by a radially
oscillating star, the best results are obtained when we use a Leapfrog
algorithm on the whole spacetime.  The reasons for this difference are
the different properties of the source term~$\Sigma^{}_{\Psi}$ and the
junction conditions~(\ref{Sfbc11}) in these two different physical
scenarios (see below for a description of the numerical implementation
of the junction conditions).  The evolution of the velocity
perturbation $\hat\beta^{(1,1)}$ is carried out by using an up-wind
scheme ($\hat\beta^{(0,1)}$ does not need to be evolved since it is
constant in time). The solution of the axial master equation~(\ref{Psi11maseq}) 
and of the conservation equation~(\ref{traseq11}) does not present 
any stability problem and the algorithm is first-order convergent. 
This is expected from the first order accurate numerical schemes used for 
the radially and differentially rotating configuration, and from the 
discontinuity of $\Psi^{(1,1)}_{,rr}$ at the stellar surface, due to 
the presence of the source terms only in the interior spacetime.

\subsection{Construction of Initial Data\label{sec:ID}}
In the following we describe how we prescribe initial data at the 
different perturbative orders: for the radial, the axial non-radial,
and the coupling perturbations.

The description of the radial perturbations requires to specify initial 
values for $\gamma^{(1,0)}\,,$ $H^{(1,0)}\,,$ $S^{(1,0)}\,,$ and $\eta^{(1,0)}$
(although these variables are not independent since they are subject to
the Hamiltonian constraint).
In this paper we consider two different types of initial conditions:
(i) We prescribe an initial profile for the fluid perturbations that 
corresponds to an eigenfunction associated with a single particular radial mode.
(ii) We prescribe initially a Gaussian profile that will excite
a broad range of normal radial modes.  

The way in which we implement both types of initial conditions consists
of setting to zero the perturbations $H^{(1,0)}\,,$ $S^{(1,0)}\,,$
$\eta^{(1,0)}$ and prescribing a non-zero initial profile for $\gamma^{(1,0)}$. 
For the first type of initial conditions (a single eigenmode) this 
implementation is equivalent to the choice of a particular time origin.
Indeed, as one can find out from equations~(\ref{eq:H10_ev})-(\ref{chi_t}) and
(\ref{eq:S10_cn}), we can choose consistently an oscillation normal mode
with eigenfrequency $\sigma^{}_n$ to have the form
\begin{eqnarray}
\gamma^{(1,0)}(t,r) & = & \gamma_n^{(1,0)}(r)\cos(\sigma^{}_n t)\,, \label{gamma10}\\
H^{(1,0)}(t,r)      & = & H_n^{(1,0)}(r)\sin(\sigma^{}_n t)\,, \\
S^{(1,0)}(t,r)      & = & S_n^{(1,0)}(r)\sin(\sigma^{}_n t)\,.
\end{eqnarray}
In this way we have that at $t=0$ only $\gamma^{(1,0)}$ is non-zero.
For the second type of initial data (Gaussian profile) our choice of
implementation is not the most general one, it just represents a
particular linear combination of eigenmodes.

To construct the first type of initial data, we can set up the eigenvalue problem 
for the radial velocity perturbation $\gamma^{(1,0)}\,$, from the
wave-type equation that can be derived  from equations (\ref{eq:H10_ev}) and
(\ref{eq:gam_t}):
\begin{equation}
\left(P(r) y^{(1,0)}_{,r}\right)^{}_{,r} + Q(r) y^{(1,0)} -
W(r) y_{,tt}^{(1,0)}=0\,, \label{eqzeta}
\end{equation}
where $y^{(1,0)} = r^2 e^{-\Lambda} \gamma^{(1,0)} =
r^2 e^{-\Phi} \xi^r_{,t}$, and $\xi^r$ is the Lagrangian displacement
of the radial oscillations.  The functions $W,P,Q$ are constructed from
quantities associated with the TOV background:
\begin{align}
& r^2 W = (\bar\rho + \bar{p})e^{3\Lambda+\Phi}\,,
\\
& r^2 P = (\bar \rho + \bar{p}) \bar{c}_s^2 e^{\Lambda+ 3\Phi}\,,
\\
& r^2 Q = (\bar\rho + \bar{p})\left[\Phi^{}_{,r}\left(\Phi^{}_{,r} -
\frac{4}{r}\right) - 8\pi\bar{p}e^{2\Lambda} \right] e^{\Lambda+3\Phi}\,.
\end{align}
By making the substitutions $y^{(1,0)} = y^{}_{0}(r) e^{i\omega t}$
and $z^{(1,0)} = P y_{,r}^{(1,0)}$ in equation~(\ref{eqzeta}) we obtain
the following pair of first-order equations
\begin{eqnarray}
y_{,r}^{(1,0)} & = & P^{-1} z^{(1,0)} \,, \label{yeq}
\\
z_{,r}^{(1,0)} & = & - (\omega^2 W + Q) y^{(1,0)}\,. \label{zeq}
\end{eqnarray}
To complete the eigenvalue problem we need to prescribe boundary conditions 
at the origin and at the surface location. At the origin, from the regularity
condition~(\ref{bcradial4}), we must have the following behaviour
\begin{equation}
y^{(1,0)}(r) = y^{}_0 r^3\,, \quad z^{(1,0)}(r) = z_0^{(1,0)}\,,
\end{equation}
which combined with equation~(\ref{yeq}) yields the following boundary
condition:
\begin{equation}
y^{}_0  = \textstyle{\frac{1}{3}}P^{-1}z^{}_0\,.
\end{equation}
%
\begin{figure}[t]
\begin{center}
\includegraphics[width=85mm, height=65mm]{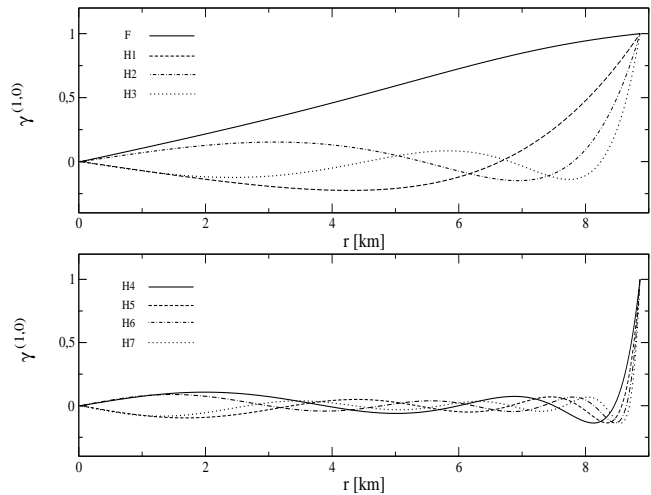}
\caption{Eigenfunctions for the radial velocity perturbation $\gamma^{(1,0)}$
corresponding to the fundamental mode and the first seven overtones.
\label{eigen_radial}}
\end{center}
\end{figure}
\begin{table}[t]
\caption{\label{tab:Rad_modes} Eigenfrequencies (in kHz) of the first eight normal modes of
    the radial perturbations.  The second column corresponds to the results from
    calculation in the frequency domain (the solution of the Sturm-Liouville problem).
    The third column corresponds to the results from a time-domain calculation, after
    applying a FFT to the resulting time series.  The fourth column shows the first 
    three normal modes computed by Kokkotas and Ruoff~\cite{Kokkotas:2000up}.}
\begin{ruledtabular}
\begin{tabular}{c | c | c | c}
Mode  & Frequency domain  & Time domain  & From~\cite{Kokkotas:2000up} \\
\hline 
    F        & 2.138              &  2.145       & 2.141        \\
    H1       & 6.862              &  6.867       & 6.871        \\
    H2       & 10.302             & 10.299       & 10.319       \\
    H3       & 13.545             & 13.590       &              \\
    H4       & 16.706             & 16.737       &              \\
    H5       & 19.823             & 19.813       &              \\
    H6       & 22.914             & 22.889       &              \\
    H7       & 25.986             & 25.964       &              \\
\end{tabular}
\end{ruledtabular}
\end{table}

The boundary condition at the surface comes from the vanishing of the 
Lagrangian pressure perturbation there.  The resulting condition reads 
[see equation~(\ref{BC10_srf})]:
\begin{equation}
(\bar\rho + \bar p)\bar{c}^2_s y^{(1,0)}_{,r}|^{}_{r=R}= 0\,.
\label{BC10_srf_Fre}
\end{equation}

The eigenvalue problem can be solved numerically by using a standard
relaxation method~(for details see~\cite{1992nrfa.book.....P}).
As discussed above, in order to obtain accurate evolutions 
of radial perturbations it is convenient to introduce the fluid tortoise
coordinate~(\ref{tort_fluid}) which provides the necessary resolution
close to the stellar surface.  The equations that one obtains for 
the eigenvalue problem in terms of the fluid tortoise coordinate have
been given in Appendix~\ref{equationstortoise}.  The eigenfrequencies and 
the associated eigenfunctions for the fundamental mode and the first seven
overtones of the radial velocity $\gamma^{(1,0)}$ are shown in 
table~\ref{tab:Rad_modes} and in Fig.~\ref{eigen_radial} respectively.
They have been normalized with respect to the absolute value of
their maximum.  Our computations of the eigenfrequencies and the associated 
eigenfunctions have second-order convergence.

In the case of the first-order axial non-radial perturbations,
we need to prescribe initial data for $\Psi^{(0,1)}$ at its time
derivative at an initial time $t=t^{}_o\,,$ say
$(\Psi^{(0,1)}_o,\partial^{}_t\Psi^{(0,1)}_o)\,$, and a certain
profile for $\hat\beta^{(0,1)}\,$, or equivalently for $\beta^{(0,1)}$
(this quantity remains contant along the evolution).
Specifying this profile is equivalent to prescribe a certain differential
rotation law for the star~\cite{Thorne:1967th,Gundlach:1999bt}.  As
we have discussed before, this law has an impact on the evolution of
$\Psi^{(1,0)}\,$, and it induces a dragging of inertial frames.  
In this work we consider two types of initial data for the 
first-order axial perturbations: (i) we prescribe a certain rotation
law $\beta^{(0,1)}\,$, and then we obtain a profile for
$\Psi^{(0,1)}_o$ from the integration of equation~(\ref{Psi01prt})
which corresponds to the particular solution $\Psi^{(0,1)}_p$.  For the
homogeneous part of $\Psi^{(1,1)}$, the dynamical one, we specify {\em zero data}:
$(\Psi^{(0,1)}_{h,o},\partial^{}_t\Psi^{(0,1)}_{h,o}) = (0,0)\,$.  (ii)
We set the star rotation to zero: $\beta^{(0,1)}=0$.  Then, we just set the 
particular solution $\Psi^{(0,1)}_p$ equal to zero and prescribe arbitrary profiles
for the homogeneous part: $(\Psi^{(0,1)}_{h,o},\partial^{}_t\Psi^{(0,1)}_{h,o})\,.$

In order to discuss the first type of initial data it is important to mention
that presently there is no sufficient knowledge on the laws of differential
rotation for neutron stars.
For nearly born neutron stars, one may obtain a description of the rotation
law from the numerical simulations of core collapse
(see, e.g.~\cite{Zwerger-Mueller1997A&A,Dimmelmeier:2002bk,Dimmelmeier:2002bm}).
In the Newtonian approach, a set of rotation laws has been introduced
motivated by mathematical simplicity and the aim of satisfying Rayleigh's
stability criterion for rotating inviscid fluids:
\begin{equation}
d(\varrho^2 \Omega)/d\varrho > 0\,,
\end{equation}
where $\varrho = r\sin\theta$ is the cylindrical radial coordinate and
$\Omega = \Omega(r,\theta)$ is the angular velocity measured by an observer at
infinity and describes the stellar differential rotation.  Among this family of
Newtonian rotation laws there is one, the {\em j-constant} law, that has been
incorporated into a General Relativistic approach~\cite{Komatsu1989MNRASA,Komatsu1989MNRASB}
by taking into consideration the dragging of the inertial frames.
In this work, we introduce a rotation law by prescribing the velocity perturbation
$\beta^{(0,1)}_{lm}$ from an expansion in (vector) spherical harmonics of the
velocity perturbation of a slowly differentially rotating star.  In the slow
rotation approximation, the fluid velocity perturbation is given by
\begin{equation}
u_{\mu}^{(0,1)} = \left( 0, u_{a}^{(0,1)} \right) =
e^{-\Phi} \left( 0,0,0, r^2 \sin^2 \theta \, \left( \Omega - \omega
\right) \right) \, , \label{u_sl_rot}
\end{equation}
where $\Omega = \Omega(r,\theta)$ is the angular velocity measured by
an observer at infinity and describes the stellar differential
rotation, while the function $\omega = \omega(r,\theta)$ describes the
dragging of inertial frames associated with the stellar rotation.  In the
case of barotropic rotating stars, the integrability condition of the
hydrostatic equilibrium equation requires that the specific angular
momentum measured by the proper time of the matter is a function of
$\Omega$ only~\cite{Komatsu1989MNRASA}, that is, $u^t u^{}_\phi =
j(\Omega)$.  In the slow rotation approximation this condition leads to the
following expression:
\begin{equation}
j^{(0,1)} (\Omega) = u^{t} u_{\phi}^{(0,1)} =
e^{-2\Phi} r^2 \sin^2 \theta \, \left( \Omega - \omega \right) \,,
\label{dj10_sp}
\end{equation}
which is a linear approximation in $\Omega$.  The choice of the
functional form of $j (\Omega)$, and hence of $j^{(0,1)} (\Omega)$,
must satisfy the Rayleigh's stability criterion against
axisymmetric disturbances for inviscid fluid:
\begin{equation}
\frac{d\,\tilde{\jmath}}{d\Omega} < 0\,,
\end{equation}
where $\tilde{\jmath}$ is the specific angular momentum
\begin{equation}
\tilde{\jmath} = (\bar\rho + \bar p) \frac{u^{}_{\phi}}{\bar\rho^{}_{0}}
\end{equation}
and $\bar\rho^{}_{0}$ is the rest mass density.  The specific angular
momentum $\tilde{\jmath}$ is locally conserved during the axisymmetric
collapse of a perfect fluid~\cite{Stergioulas:2003ep}.  A common
choice for $j(\Omega)$ that satisfies these
conditions~\cite{Komatsu1989MNRASA, Komatsu1989MNRASB} is the
following:
\begin{equation}
j^{(0,1)}(\Omega) = A^2 \left( \Omega^{}_c - \Omega \right)\,,
\label{jOmega}  
\end{equation}
where $\Omega^{}_{c}$ is the angular velocity at the rotation axis
and $A$ is a constant that describes the amount
of differential rotation.   We can now find the rotation law from 
equations~(\ref{dj10_sp}) and~(\ref{jOmega}).  The result is:
\begin{equation}
\Omega(r,\theta) = \frac{A^2\Omega^{}_{c} + e^{-2\Phi} r^2 \sin^2\theta\,
\omega(r,\theta)}{A^2 + e^{-2\Phi} r^2 \sin^2\theta} \,. \label{j-cons}
\end{equation}
In the Newtonian limit this equation reduces to the $j$-constant
rotation law used in Newtonian analysis~\cite{1986ApJS...61..479H}:
\begin{equation}
\Omega(r,\theta) = \frac{A^2\Omega^{}_c}{A^2 + r^2 \sin^2 \theta}
\,. \label{j-cons_New}
\end{equation}
A uniform rotating configuration with $\Omega = \Omega^{}_c$ is attained
for high values of $A$ ($A \rightarrow \infty$). On the
other hand, for small values of $A$, the law (\ref{j-cons}) describes,
in the Newtonian limit, a configuration with constant angular momentum.

The only non-vanishing vector spherical harmonic components of $u_{a}^{(0,1)}$
are given by the expansion
\begin{equation}
u_{\phi}^{(0,1)} = \sum^{}_{l,m} \beta_{lm}^{(0,1)} S_{\phi}^{lm}  \,,
\end{equation}
where $\beta^{(0,1)}_{lm}$ are the velocity perturbations (which depend on
the coordinate $r$) that were introduced in equation~(\ref{deltau})
[we have now restored the spherical harmonic indices $(l,m)$].
The different $\beta_{lm}^{(0,1)}$ are obtained by using the properties
of the inner product among the different elements of the basis of axial vector 
spherical harmonics.  Their expression is
\begin{equation}
\beta_{lm}^{(0,1)} = \frac{1}{\sqrt{l(l + 1)}} \,
\int^{}_{S^2} \sin\theta d\theta d\phi \, u_a^{(0,1)}
S_{b}^{lm} \gamma^{ab} \,, 
\label{innpr}
\end{equation}
where $\gamma^{ab}$ is the contravariant metric on $S^2$.

In order to determine the initial profile for the axial velocity perturbations
we just have to introduce the rotation law~(\ref{j-cons}) into the form
of the velocity perturbations~(\ref{u_sl_rot}).  We obtain
\begin{equation}
u_{\phi}^{(0,1)} =  \frac{e^{-\Phi} A^2}{A^2 + e^{-2\Phi} r^2 \sin^2 \theta }
\left(r^2\sin^2\theta\,\Omega^{}_{c} + \sum^{}_{lm} k_0^{lm} S_{\phi}^{lm}\right)\,,
\label{betjlaw}
\end{equation}
where we have used equation~(\ref{frmdr-rel}), which relates the metric
perturbations with the frame dragging function $\omega\,.$ It is important
to notice that the first term in~(\ref{betjlaw}) corresponds to the
the Newtonian j-rotation law (up to the factor $e^{-\Phi}$), to which we
will refer as the {\em nearly Newtonian j-rotation law}.   The other
terms account for the dragging of the inertial frames.
To obtain the velocity perturbations $\beta^{(0,1)}_{lm}$ from~(\ref{innpr})
we just need to introduce there expression~(\ref{betjlaw}). In doing
this, we can easily determine the expression for the nearly
Newtonian term, but we need to pay special attention to the relativistic 
corrections.  The difficulty arises from the expansion in axial vector
harmonics in~(\ref{betjlaw}) containing the unknowns, $k_0^{lm}$,
 of the differential equation~(\ref{k0-fradr}).  Due to the form
of (\ref{betjlaw}), the inner product in equation~(\ref{innpr}) will
contain products of harmonics with different harmonic indices.
In order to decouple these terms in the relativistic corrections we assume that
the dominant contributions are provided by the metric perturbations $k_0^{lm}$
that have the same harmonic indices as in the nearly Newtonian rotation law.
For the nearly Newtonian part, we have found solutions only for $A > e^{-
  \Phi(R^{}_s)} R^{}_{s}$.  From this relation and from the values assumed by the
metric function $\Phi$ in the stellar model considered in this work (see
subsection~\ref{background}) we can give the following estimation for the allowed
values of $A$:
\begin{equation}
 \frac{1}{2}  < \frac{A^2}{A^2 + e^{-2 \Phi} r^2 \sin^2 \theta} \le 1 \,.
\end{equation}
Then, using these inequalities, the approximation that we obtain for
the relativistic terms of equation~(\ref{betjlaw}) is given by:
\begin{equation}
u_{\phi}^{(0,1)} = \frac{ A^2 r^2 \sin^2 \theta \,  e^{-\Phi} \,  {}
 }{A^2 + e^{-2 \Phi} r^2 \sin^2 \theta }  \, \Omega^{}_c + \alpha^{}_0  e^{-\Phi}
 \sum^{}_{lm} k_0^{lm} S_{\phi}^{lm}  \, , \label{betjlawred}
\end{equation}
where here $\alpha^{}_0$ is a constant such that $\alpha^{}_0 \in (0.5,1]$.
The coefficients of the expansion of this equation in axial vector harmonics 
have the following form 
\begin{equation}
u_{\phi,l0}^{(0,1)} = \left\{ \begin{array}{ll}
\alpha^{}_0 e^{-\Phi} k_0^{l0}S_{\phi}^{l0} & \quad \textrm{for}~l~\textrm{even} \,,
   \\ \nonumber\\ 
\beta^{(0,1)}_{l0} S_{\phi}^{l0} & \quad \textrm{for}~l~\textrm{odd} \,,
\end{array} \right.   
\end{equation}
where the coefficients of the components with odd $l$ are given by the 
following expression:
\begin{equation}
\beta^{(0,1)}_{l0} = \frac{e^{\Phi}\Omega^{}_c}{\sqrt{l(l+1)}}f^{}_{l0}\left(x,A\right)
 + \alpha^{}_0 e^{-\Phi}\, k_0^{l0}\,, \label{bet01_IC}
\end{equation}
where $x=r e^{-\Phi}$~(not to be confused with the fluid tortoise coordinate).
The functions $f^{}_{l0}(x,A)$ for $l=1$ and $l=3$ are
given by the following expressions
\begin{widetext}
\begin{eqnarray}
f^{(0,1)}_{10} & = & - 4.342 A^2 \left[1-\frac{A^2}{x \sqrt{A^2 + x^2} }
\ln\left(\frac{\sqrt{\sqrt{A^2 + x^2}+x}}{\sqrt{\sqrt{A^2 + x^2}-x}}\right)
\right] \,, \label{bet01_j_law_l1} \\
f^{(0,1)}_{30} & = & - 2.708 A^2 \left[  1 + 7.5 \frac{A^2}{x^2}
- \frac{6 A^2}{x\sqrt{A^2 + x^2}}\left(1+\frac{5}{4}\frac{A^2}{x^2}\right)
\ln\left(\frac{\sqrt{\sqrt{A^2 + x^2}+x}}{\sqrt{\sqrt{A^2 + x^2} - x}}\right)
\right] \,. \label{bet03_j_law_l1}
\end{eqnarray}
\end{widetext}
which have been derived by imposing the condition $A > R^{}_s e^{-\Phi(R^{}_s)}$.

\begin{figure}[t]
\begin{center}
\includegraphics[width=85mm, height=65mm]{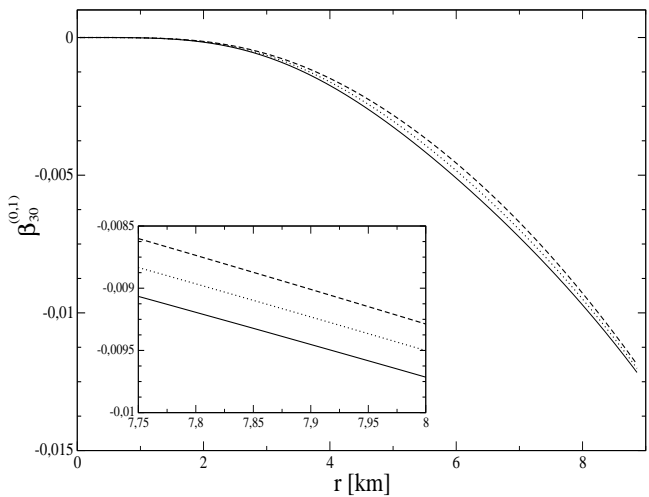}
\caption{\label{fig:beta01}Profiles of the $l=3$ component of the axial velocity perturbation 
$\beta^{(0,1)}_{30}$ determined for a $j$-rotation law with $A=10$ km and a rotation period at 
the axis of $T=10$ ms. The solid line corresponds to the axial velocity associated with the 
nearly Newtonian $j$-law, while the dotted and dashed lines correspond to the velocity perturbations 
in~(\ref{bet01_IC}) with $\alpha = 0.5$ and $\alpha = 1$ respectively.}
\end{center}
\end{figure}

In the limit of $A \rightarrow \infty$, the functions $f^{}_{l0}$ behave
as expected, that is, the nearly Newtonian part becomes:
\begin{equation}
\lim^{}_{A \rightarrow \infty} \frac{e^{\Phi}\Omega^{}_c}{\sqrt{l(l+1)}}
f^{}_{l0} S_{\phi}^{l0} = \left\{
\begin{array}{ll} 
e^{-\Phi}\Omega^{}_c  r^2 \sin^2 \theta & \quad  \textrm{for}~ l=1 \,,
\\ \nonumber\\ 
0 & \quad \textrm{for}~ l \geq 3 \,,
\end{array} \right.
\end{equation}
which corresponds to a uniformly rotating configuration, where $\Omega =
\Omega^{}_c$ is the angular velocity as measured at infinity.
Since in this work we are mainly interested in the impact of the
non-linear coupling of the stellar oscillations in the gravitational
wave emission, we will consider only the case $l=3$, as the dipolar 
term $l=1$ does not produce gravitational waves.

In Fig.~\ref{fig:beta01} we plot the profiles for the axial velocity
perturbation $\beta_{30}^{(0,1)}$.  It shows that the two solutions
corresponding to the extreme values of the constant $\alpha^{}_0$ disagree
in an amount below $10\,\%$.
In the following simulations we will use for
simplicity only the nearly Newtonian term by obtaining results which
are correct to better than $10\,\%$.

With regard to second choice of initial conditions for the axial non-radial
perturbations, namely $\beta^{(0,1)}=0\,$, it is going to be used in this
paper to investigate how the second-order metric perturbations corresponding
to the coupling terms can be affected by the radial pulsations of the star.
Actually, it is well known that the axial components of the gravitational-wave signal
emitted by a non-rotating compact star can only contain
the imprint of the spacetime $w$-modes. The question we can ask is then
how this particular signal looks like when it couples with the radial
pulsations of the star.
What we do in practice is to generate $w$-modes at first-order
and to study the corresponding signal at second-order (in the sector of
the coupling of radial and non-radial modes) by looking at its dependence
on the radial pulsations of the star.  The $w$-mode oscillation is excited
by the standard procedure of scattering a (narrow) Gaussian pulse of 
gravitational waves on the star.

For the non-linear coupling perturbations we have to prescribe initial
data for $\Psi^{(1,1)}$, $\Psi^{(1,1)}_{,t}$ and $\beta^{(1,1)}$ at
an initial time $t=t^{}_o$. Since we are mostly interested in the non-linear
effects generated from the coupling between linear perturbations and
not in the evolution {\em per se} of second-order perturbations,
we set these quantities  initially  to zero.

To sum up, combining the different types of initial data for the
first-order perturbations we can study the following two different
scenarios: (i) the scattering of a gravitational wave by a radially
oscillating star. (ii) A radially pulsating star with with
differential rotation.

\subsection{Boundary conditions\label{sec:BD}}
The behaviour at the stellar center of the radial, axial non-radial, and non-linear
coupling perturbations is given in equations~(\ref{bcradial1})-(\ref{bcradial4}) and
(\ref{BC01_orig}).  In order to deal numerically with the centrifugal term of
the potential of the axial master equation, $l(l+1)r^{-2}$, we use a numerical grid
where the first grid-point, $r^{}_1\,$, does not coincide with the center $r=0$ but
instead is located at $r^{}_1 = \Delta r\,$, being $\Delta r$ the grid spacing.
Then, we impose at $r^{}_1$ the regularity conditions
(\ref{bcradial1})-(\ref{bcradial4}) and (\ref{BC01_orig}).
For instance, in the case of the variable $\Psi^{(0,1)}$, which behaves as
$\Psi^{}_0 (t) r^{l+1}$ [equation~(\ref{BC01_orig})],  we assume that this behaviour is
in a good approximation also valid at the grid points $r^{}_1$ and $r^{}_2$, and then we
determine the value of $\Psi^{(0,1)}$ at $r^{}_1$ from the following relation:
\begin{equation}
\Psi^{(0,1)} (r^{}_1) = \Psi^{(0,1)} (r^{}_2) \left(\frac{r^{}_1}{r^{}_2}\right)^{l+1} \,.
\end{equation}
Let us consider now the boundary conditions at the stellar surface, where the
pressure vanishes.  In this way, the unperturbed stellar surface is
$\bar\Sigma = \{x ~|~ r(x)=R^{}_s\}$ where $R^{}_s$ is the stellar radius of the TOV model,
where the background pressure $\bar p$ vanishes.
Due to the radial pulsations,  the surface of the perturbed star
does not coincide with the static surface, but is given by
$\Sigma\equiv\{ x + \lambda\xi^{(1,0)} ~|~ x\in\bar\Sigma\}\,$,
where $\xi^{(1,0)}$ is the Lagrangian displacement vector of a fluid element.
The axial non-radial modes only induce rotation of the star and hence,
they do not change the shape of the stellar surface.
Therefore, at first order we have to impose
the vanishing of the total pressure on the perturbed surface $\Sigma$ for the
radial perturbations,
which is equivalent to the relation (\ref{BC10_srf}), and the continuity conditions
on the static surface for the axial non-radial perturbations.

The radial perturbations are evolved using the {\em x-grid} (the one
determined by the fluid tortoise coordinate), then we interpolate the
solution to the {\em r-grid} (the one determined by the area radial coordinate).
In terms of the fluid tortoise coordinate $x$, the surface condition
(\ref{BC10_srf}) is given in equation~(\ref{bcx}) of
Appendix~\ref{equationstortoise}, with $R^{}_x$ being the value of the
fluid tortoise coordinate corresponding to the static stellar radius
$R^{}_s$.  This condition will be certainly satisfied by finite values
for $\gamma^{(1,0)}$ and $\gamma^{(1,0)}_{,x}$, since pressure,
density and speed of sound go to zero at the surface for a polytropic
equation of state.  We then determine, at every time step, the finite
value of $\gamma^{(1,0)}$ on the surface by using a second-order
polynomial extrapolation.  The surface boundary conditions for the
other three radial variables are then directly determined from the
perturbative equations.

The solution of the axial master equation (\label(\ref{Psi01maseq}) is
decomposed into two pieces: an homogeneous part $\Psi_h^{(0,1)}$ that satisfies
the homogeneous equation, and a static particular solution $\Psi_p^{(0,1)}$
that satisfies the equation~(\ref{Psi01prt}).
These equations are solved numerically on the whole spatial grid without imposing
junction conditions.  In this way, the continuity of $\Psi_h^{(0,1)}$ and $\Psi_p^{(0,1)}$
and its first spatial derivatives, required by the junction conditions described in 
Section~\ref{axialpert}, follows automatically.

The junction conditions for the non-linear perturbations require some
approximations (see~\cite{Sperhake:2001si,Sperhake:2001xi} for a treatment
of the junction conditions in a similar non-linear context).
In an Eulerian gauge the perturbed surface will not coincide in general with the
surface of the background equilibrium configuration.
As a consequence, some perturbative quantities may take unphysical values near
the surface during the expansions and contraction phases of the star.
For instance, while the background enthalpy and density go to zero as we approach
the static stellar surface, the associate radial perturbations have an oscillatory
character with amplitudes that, near the surface, are bigger than the background values.
Therefore, in a contraction phase, when the Eulerian radial
perturbations of enthalpy and density are negative, the total
mass-energy $\bar\rho + \delta\rho^{(1,0)}$ may take negative values.
Furthermore, the low densities which are present at the
outermost layers of the star may also produce some numerical errors in the
simulations~\cite{2003PhRvD..68b4002H}.  These problems can be avoided 
if we do not impose the matching conditions for the non-linear perturbations at
the static stellar surface, as we do for the linear perturbations, but
at a hypersurface that during the evolution remains always slightly inside the
unperturbed star~\cite{Sperhake:2001si,2003PhRvD..68b4002H}.
In practice this procedure is implemented as follows:
(i) we estimate the amplitude of the surface movement due to radial
pulsations. (ii) We compute the linear (radial and non-radial) perturbations up to
the static surface $R^{}_s$. (iii) We impose the junction conditions for the non-linear
perturbations on a hypersurface that is always inside (even during the contraction phase)
the unperturbed star. (iv) We estimate the effects on the gravitational signal.
In this sense, it is important to mention that in doing this approximation
we are removing the outer layers of the star, where the density is very low,
which in practice means that we are neglecting less than one percent of the stellar
mass, which does not produce significative changes in the waveforms and spectra of the
gravitational signal.   Moreover, by using an appropriate choice of initial conditions,
the movement of the perturbed surface due to radial pulsations can be very small
$\sim 10^{-3} R^{}_s$. Therefore, this approximative procedure leads to neglect only one or
two grid points near the stellar surface.  However, on the negative side, since the source
term for the axial master equation~(\ref{Psi11maseq}) takes its maximum
amplitude at the stellar surface, this removal of one or two grid points induces
an error of about $5\,\%$ in the gravitational wave signal.

The procedure we have just described allows us to simplify the junction
condition~(\ref{Sfbc11}) for the two initial configurations that we consider in this
paper.   In the case of the scattering of an axial gravitational wave by a
radially pulsating star, the velocity perturbations $\hat\beta^{(0,1)}$ and
$\hat\beta^{(1,1)}$ are set to zero.  Therefore, by using the continuity of the first 
order perturbations, the junction condition~(\ref{Sfbc11}) reduces to the continuity of
$\Psi^{(1,1)}_{,r}$.  We then evolve the axial master equation (\ref{Psi11maseq}) for
$\Psi^{(1,1)}$ on the whole numerical grid, with the source terms only present in 
the interior of the star.  In this way the junction conditions on the matching surface 
are automatically satisfied.  For the second situation that we study, the non-linear
coupling  in a radially pulsating and differentially rotating star, we can only use
the continuity of the linear perturbations to reduce~(\ref{Sfbc11}) to the following
expression:
\begin{equation}
e^{-\Lambda}  \Psi^{(1,1)}_{,r} + 16\pi r\hat\beta^{(1,1)}\,.
\end{equation}
In this case, the numerical procedure to solve the axial master
equation~(\ref{Psi11maseq}) goes as follows: (i) In the
interior spacetime we implement an up-wind scheme. (ii) At the
matching hypersurface, the junction conditions provide the values of
the axial function $\Psi^{(1,1)}$ and its time and spatial derivatives.
(iii) Finally, the values obtained at the matching hypersurface 
are used to evolve the axial master function $\Psi^{(1,1)}$ in the
exterior (the $(1,1)$ Regge-Wheeler equation) by using a Leapfrog scheme.

\begin{figure}[t]
\begin{center}
\includegraphics[width=85mm]{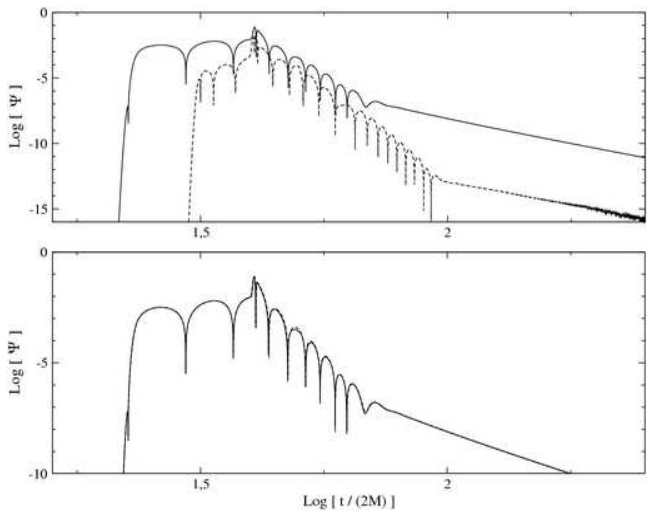}
\caption{\label{fig:Cou_ScaF} Waveforms (in logarithmic scale) from the scattering of
an $l=2$ axial gravitational wave from a star oscillating radially in the fundamental 
mode (see Table~\ref{tab:Rad_modes}).
The top panel presents the first-order axial non-radial master function
$\Psi^{(0,1)}$ (solid line) and the second-order coupling function
$\Psi^{(1,1)}$ (dashed line).  The bottom panel exhibits the waveform $\Psi^{(0,1)}$ 
(solid line) together with the total signal $\Psi^{(0,1)} + \Psi^{(1,1)}$ (dashed line).}
\end{center}
\end{figure}

\section{Results }
\label{results}

The following section is devoted to discuss the results of time-domain numerical 
simulations describing the non-linear coupling of radial and axial non-radial 
oscillations of a static star in two different physical settings: 
(i) The scattering of a gravitational wave by a radially pulsating star
(subsection~\ref{scenario1}).
(ii) A differentially rotating and radially oscillating star
(subsection~\ref{scenario2}).

\subsection{Effects of radial pulsations on the scattering of a gravitational wave}
\label{scenario1}
The aim of these simulations is to investigate the possible signature of the
radial oscillations of the star in scattered gravitational waves.
The initial first order axial perturbation correspond to an $l=2$ mode
with vanishing velocity perturbation $\beta^{(0,1)}$.  The profile of
the axial master function $\Psi^{(0,1)}$ is a Gaussian pulse 
\begin{equation}
\Psi^{(0,1)}(t^{}_o,r) = A^{(0,1)} e^{-q(r-r_{0})^2} \, , 
\label{Imp_psi01}
\end{equation}
centered at $r_0 = 20$ km with amplitude $A^{(0,1)} = 0.1$ and width
parameter $q=1.25$. The star oscillates in one of the radial eigenmodes 
corresponding to the frequencies of Table~\ref{tab:Rad_modes}.  

\begin{figure}[t]
\begin{center}
\includegraphics[width=85mm]{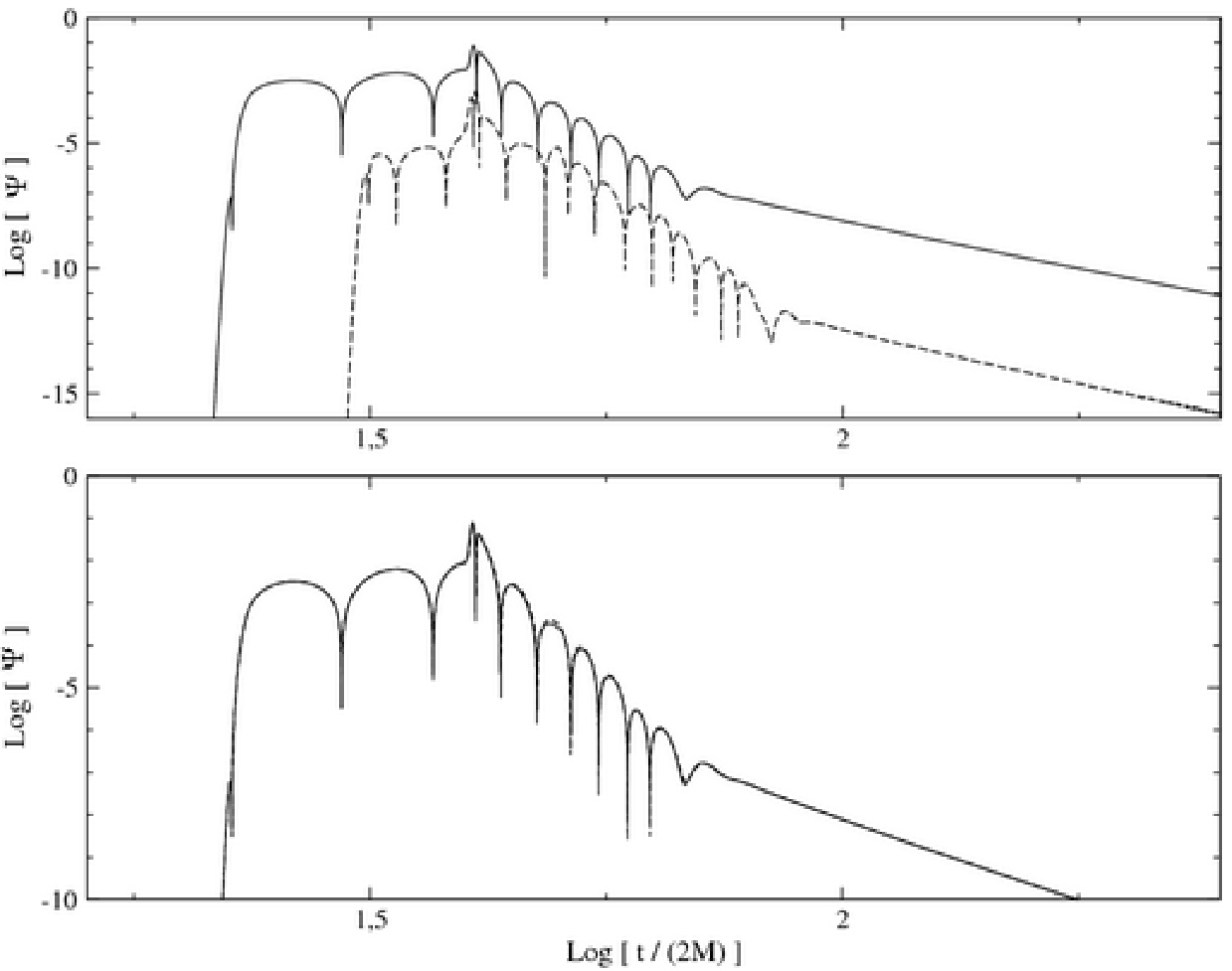}
\caption{\label{fig:Cou_ScaH1} Waveforms (in logarithmic scale) from the scattering of
an $l=2$ axial gravitational wave from a star oscillating radially in the first overtone
H1 (see Table~\ref{tab:Rad_modes}).
The top panel presents he first-order axial non-radial master function
$\Psi^{(0,1)}$ (solid line) and the second-order coupling function
$\Psi^{(1,1)}$ (dashed line). The bottom panel exhibits the waveform $\Psi^{(0,1)}$ 
(solid line) together with the total signal $\Psi^{(0,1)} + \Psi^{(1,1)}$ (dashed line). }
\end{center}
\end{figure}

In Figs.~\ref{fig:Cou_ScaF} and~\ref{fig:Cou_ScaH1}, we present the $l=2$ waveforms 
associated with the first-order axial non-radial master function $\Psi^{(0,1)}$ 
and the master function $\Psi^{(1,1)}$ describing the coupling.  We find that 
the correction to the signal coming from the coupling is less than $2\,\%$ when 
the radial pulsations are excited by the F-modes, and less than $0.1\,\%$ for 
higher overtones. These corrections do not modify the properties of the waveforms 
and spectra associated with the first-order perturbations.  We find similar 
results for higher overtones and their combinations. In summary, in this particular
scenario we have not found any significant amplification of the gravitational
wave signal due to the coupling of the radial and axial non-radial first-order
perturbations.

\subsection{Coupling between radial pulsations and differential rotation}
\label{scenario2}

The aim of the simulations we present in this subsection is to investigate
the effect of the coupling of radial oscillations with differential rotation,
as described by axial non-radial first-order perturbations, in the gravitational
wave signal. The axial rotation of the fluid is described with a good approximation
by the nearly Newtonian $j$-constant rotation law described in the
subsection~\ref{sec:ID}.  All the simulations in this subsection have a first-order
rate of convergence and are long term stable.

\subsubsection{Single radial mode oscillation}

We start with simulations in which we consider single mode excitation of 
the radial oscillations, in particular the first overtone H1.
The initial velocity perturbation $\gamma^{(1,0)}$ is taken to be 
[see equation~(\ref{gamma10})] 
\begin{equation}
\gamma^{(1,0)} = G^{(1,0)}\gamma^{(1,0)}_n \,, \label{iniradialmode}
\end{equation}
with $G^{(1,0)}=0.001$. For this amplitude we can be sure that the region of 
the spacetime that the motion of the surface covers is confined to a narrow 
slice around the static equilibrium configuration. Therefore, the issues 
related to negative values of the mass-energy density in an Eulerian description 
can be solved approximatively as we explained above.

The values of the differential rotation parameter $A$ and the angular velocity at the
rotation axis in these simulations are $A = 15$ km and $\Omega^{}_c = 2.09\times 10^{-3}$ km$^{-1}$.
This latter (corresponding tp a period of $10$ ms) is small compared
with the mass shedding limit value $\Omega^{}_K$: $\Omega^{}_c/ \Omega^{}_{K} =
6.45\times 10^{-2}$. The profile of the axial velocity perturbation
$\beta^{(0,1)}$ for $l=3$ is given in equation~(\ref{bet03_j_law_l1}) and 
depicted in Fig.~\ref{fig:beta01}.  For this initial configuration of the 
differential rotation, the coupling non-linear perturbations are dominated by 
the harmonic $l=3\,$.

\begin{figure}[t]
\begin{center}
\includegraphics[width=85mm,height=65mm]{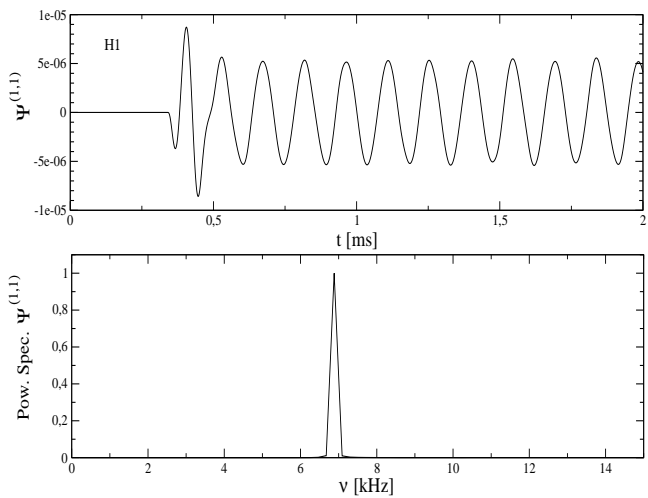}
\caption{\label{fig:Coupl_Psi11_H1} Coupling between radial mode oscillations (overtone H1) 
and differential rotation (with rotation period $T=10$ ms). 
Waveform $\Psi^{(1,1)}$ for $l=3$ multipole ({\it top panel}). 
A (transient) excitation of the first $l=3$ $w$-mode at early times is present, 
followed then by a periodic oscillation driven by the radial mode. The Fourier
spectrum of the periodic oscillation ({\it bottom panel}) is dominated by the
frequency $\nu_{\rm H1}\simeq 6.8$ kHz corresponding to the H1 radial pulsation. }
\end{center}
\end{figure}

In Fig.~\ref{fig:Coupl_Psi11_H1} we show the $l=3$ waveform ({\it top panel}) 
extracted at $100$ km apart from the stellar center: after an initial transient 
phase dominated by $w$-mode ringing (see below), the $\Psi^{(1,1)}$ waveform exhibits 
a monocromatic oscillation at the frequency of the H1 mode; i.e., $\nu_{\rm H_1}\approx 6.8$ kHz,
as confirmed by the Fourier spectrum of the signal ({\it bottom panel}). 
The presence of the $w$-mode in the initial phase of the waveform is related to the 
choice of initial data that we have done for simplicity; 
i.e., $\Psi^{(1,1)}=\Psi^{(1,1)}_{,t}=\beta^{(1,1)}=0$. This is, in general, non consistent, 
because these three quantities are not independent; indeed, a correct set up
of the initial data for $\Psi^{(1,1)}$ would require the solution of the ``coupling'' axial 
constraints (not reported here) in a way similar to what is usually done for the first-order 
polar and axial perturbations (see for example Ref.~\cite{Nagar:2005ea}). If this is not done, 
the system requires to relax to the ``correct'' solution through the release of a GWs burst
of $w$-modes. However, since our interest is focused here on the stationary phase driven by 
the radial oscillations of the star, we decided to avoid the complications related to correctly 
capture the initial transient.

\subsubsection{Amplification effects}

\begin{figure}[t]
\begin{center}
\includegraphics[width=85mm,  height=75mm]{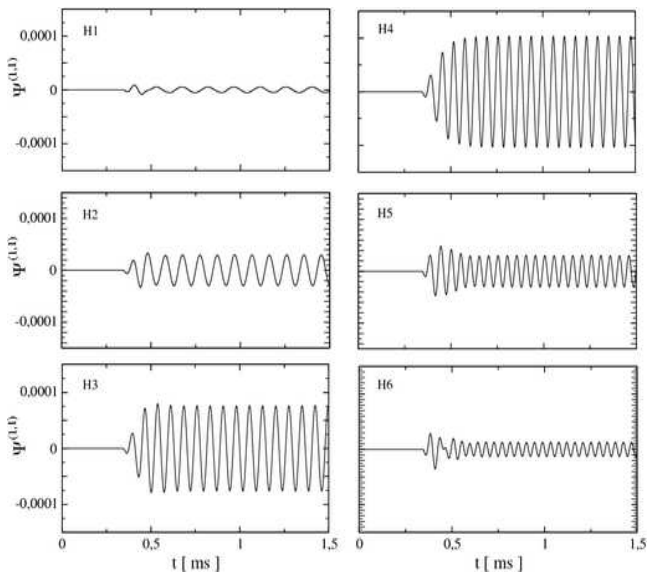}
\caption{\label{fig:Coupl_Psi11_H16} Comparison of six $\Psi^{(1,1)}$ waveforms for the
$l=3$ multipole. The differential rotation law is the same as the one used in
Fig.~\ref{fig:Coupl_Psi11_H1}.
The radial pulsations considered correspond to single mode oscillation from H1 to H6 
overtone. These plots show that a resonance effect take place in $\Psi^{(1,1)}$.
See text for a discussion.}
\end{center}
\end{figure}

\begin{figure}[t]
\begin{center}
\includegraphics[width=85mm, height=75mm]{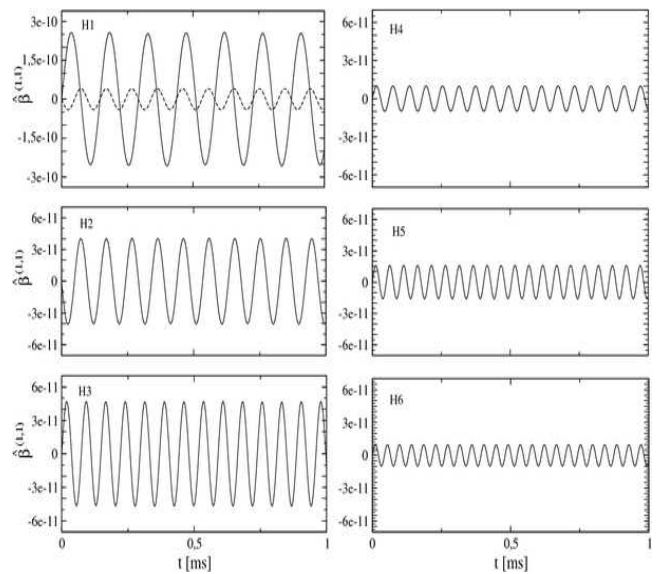}
\caption{\label{fig:Coupl_bre11} Time evolutions (the same runs as in 
Fig.~\ref{fig:Coupl_Psi11_H1}) of the second-order fluid velocity perturbation 
$\hat\beta^{(1,1)}\,$. The perturbation $\hat \beta^{(1,1)}$ has been averaged 
over the star at each time step and plotted here. The dashed line in the left top 
panel (H1) reports the perturbation for the H2 eigenmode and is superposed for 
comparison. }
\end{center}
\end{figure}

In this section we extend the previous analysis to simulations where
the radial pulsations are excited using combinations of the
fundamental mode and higher overtones.  In general, the waveforms and
spectra present the same features as the H1 case described above.
However, an interesting amplification is observed in simulations where
the radial oscillations contain a frequency close to the one of the
axial $w$-mode. In Fig.~\ref{fig:Coupl_Psi11_H16}, we show the axial
master function $\Psi^{(1,1)}$ for six runs using different radial
overtones.  It is clear there that $\Psi^{(1,1)}$ increases in
amplitude as we increase the order of the radial normal mode from the
first to the fourth overtone, while it decreases as we go to the fifth
and sixth overtones.  The amplitude of $\Psi^{(1,1)}$ for the H4 case
is about sixteen times that of $\Psi^{(1,1)}$ for the H1 case. For
this stellar model, the $l=3$ spacetime mode has frequency $16.092$
kHz which is between the frequencies of the third and fourth radial
overtones, $13.545$ kHz and $16.706$ kHz respectively.  Moreover, this
effect takes place despite the energy and the maximum displacement of
the surface of the radial modes decrease proportionally to the order
of the radial modes~(see Table~\ref{tab:damping_time}).  The
interpretation of this amplification as a sort of resonance effect is
supported by the structure of the axial master
equation~(\ref{Psi11maseq}) and its analogy with a forced oscillator.
The amplification arises when one of the {\em natural} frequencies,
determined by the form of the axial potential $V$ [see
equations~(\ref{rwpotential}, \ref{Psi11maseq})], is sufficiently close to
the frequency (or frequencies) associated with the {\em external}
force.  In the case when the non-radial perturbations are static and
just describe differential rotation, the frequencies corresponding to
the {\em external} force are determined by the structure of the radial
oscillations.  This picture is confirmed by the fact that this
amplification effect disappears when the axial potential is
arbitrarily removed.  In addition, the fluid velocity perturbation
$\hat\beta^{(1,1)}$, which satisfies a completely different
equation~(\ref{traseq11}), does not show any such amplification and
instead decreases proportionally with the order of the radial mode
(see Fig.~\ref{fig:Coupl_bre11}).

\begin{table*}
\caption{\label{tab:damping_time} Quantities associated with radial normal modes
and their coupling to the first-order axial differential rotation: Energy, $E^{(1,0)}_{n}$,
and maximum stellar surface displacement $\xi^{(1,0)}_{\rm sf}$ of the radial eigenmodes
[From initial conditions~(\ref{iniradialmode}) with $G^{(1,0)}=0.001$];
power, $\dot E^{(1,1)}_{30}$,
emitted in gravitational waves to infinity from the coupling between the radial eigenmode 
and the axial differential rotation; estimated values of the damping times, $\tau^{(1,1)}_{30}\,;$
and number of oscillation periods, $N^{}_{\rm osc}\,,$ that takes for
the non-linear oscillations to radiate the total energy initially
contained in the radial modes.}
\begin{ruledtabular}
\begin{tabular}{l c c c c c c c}
~Normal Mode:~ & F & H1 & H2 & H3 & H4 & H5 & H6 \\
\hline \hline\\[-3mm]
$E^{(1,0)}_{n}~[10^{-8}$ km]         & $35.9$               & $4.2$               & $1.37$   & $0.62$   & $0.34$    & $0.21$  & $0.14$   \\
$\xi^{(1,0)}_{\rm sf}$ [m]           & $12.65$              & $4.02$              & $2.66$   & $2.02$   & $1.64$    & $1.38$  & $1.19$   \\
$<\dot E^{(1,1)}_{30}>~[10^{-13}]$  & $1.85\times 10^{-6}$ & $6.83\times 10^{-2}$ & $4.85$   & $56.03$  & $157.01$  & $19.08$ & $4.45$   \\
$\tau^{(1,1)}_{30}$ [ms]             & $6.49\times 10^{9}$  & $20.49\times 10^{3}$ & $94.15$  & $3.69$   & $0.72$    & $3.67$  & $10.51$  \\
$N^{}_{\rm osc}$                     & $1.39\times 10^{10}$ & $1.408\times 10^{5}$ & $971.58$ & $49.99$  & $12.07$   & $72.78$ & $240.77$ \\
\end{tabular}
\end{ruledtabular}
\end{table*}

\begin{figure}[t]
\begin{center}
\includegraphics[width=85mm, height=65mm]{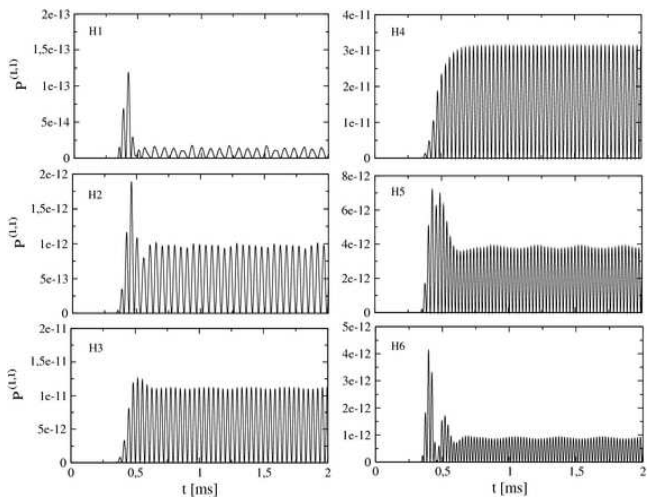}
\caption{\label{fig:Coupl_Power11_H1H6}  Emitted power in gravitational waves 
$P^{(1,1)} = \dot{E}_{30}^{(1,1)}$ for the six runs of Fig.~\ref{fig:Coupl_Psi11_H16}.}
\end{center}
\end{figure}

In agreement with these results, the power emitted in gravitational waves to infinity 
exhibits the same amplification too (see Fig.~\ref{fig:Coupl_Power11_H1H6}).
There are obviously two different contributions to the power, one due to the $w$-mode
excitation related to the initial data choice and the second related to the periodic
emission driven by the stellar radial pulsations. The relative strength of these two
contributions to the power changes as we change the radial eigenmode. The periodic 
emission increases for the first eigenmodes and reaches its maximum for the H4 overtone;
it decreases then for H5 and H6.

\begin{figure}[t]
\begin{center}
\includegraphics[width=85mm, height=65mm]{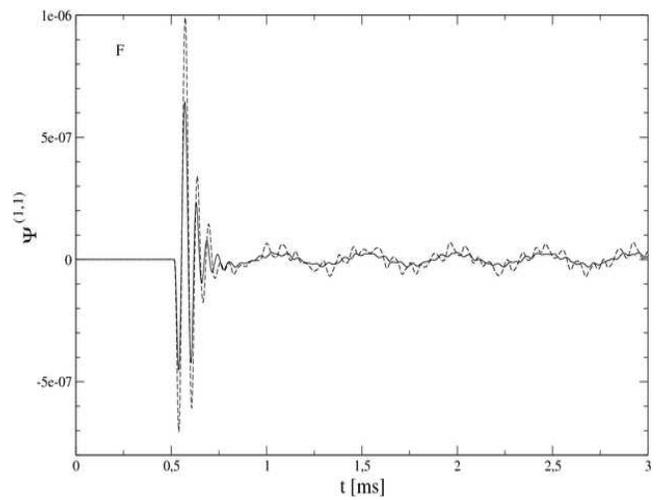}
\caption{\label{fig:Coupl_Psi11_FF} Waveforms generated by the coupling between 
axial differential rotation and the radial fundamental mode. We show superposed
the waveform with the matching condition at $r =8.64$ km (dashed line) and  
at $r=7.75$ km (solid line). See text for a discussion.}
\end{center}
\end{figure}

Let us now look at the coupling between differential rotation and
radial pulsations in the fundamental mode F.  The $\Psi^{(1,1)}$
waveform is reported in Fig.~\ref{fig:Coupl_Psi11_FF}. After the
initial $w$-mode transient, the signal is dominated by a periodic
pulsation at the radial F-mode frequency. We notice that the 
amplitude of the oscillation is smaller than in the previous cases
due probably to the fact that the radial F-mode frequency is not 
enough close to the first $w$-mode frequency. The periodic part of the 
signal exhibits high frequency oscillations associated with the low 
density region near the stellar surface: in Fig.~\ref{fig:Coupl_Psi11_FF} 
we depict (with a dashed line) the waveform obtained when the junction
conditions have been imposed at a radial coordinate $r=8.64$ km, 
(which is equivalent to neglect $0.3\,\%$ of the stellar mass) and
with a solid line the case when the matching surface stands at 
$r=7.75$ km (corresponding to neglecting a  $6.3\,\%$ of the
total mass). In the later case the numerical noise is reduced
and the waveform looks closer to the expected form.

Our perturbative approach does not include backreaction
(see, e.g.~\cite{Balbinski:85be,Yoshida:95ye}); in other words,
it does not account for the damping of the radial oscillations or the slowing down 
of the stellar rotation due to contribution of the non-linear coupling to the 
energy and angular momentum loss in gravitational waves. Backreaction could be
studied by looking at higher perturbative orders.   Nevertheless,
we can provide a rough estimate of the damping time of the radial pulsations
by assuming that the energy emitted is completely supplied by the first-order 
radial oscillations, and that the power radiated in gravitational waves is 
constant in time.  In this way, the damping time is given by the following 
expression:
\begin{equation}
\tau^{(1,1)}_{lm} \equiv \frac{E^{(1,0)}_n}{<\dot{E}^{(1,1)}_{lm}>}\,,
\label{dam_time}
\end{equation}
where $E^{(1,0)}_n$ is the energy of a radial eigenmode~(see
Table~\ref{tab:damping_time}), and $<\dot{E}^{(1,1)}_{lm}>$ is the averaged
value of the non-linear coupling contribution to the power emitted.
The results for $\tau^{(1,1)}_{30}$ are shown in Table~\ref{tab:damping_time}.
Moreover, in the last row of Table~\ref{tab:damping_time} we give an estimation
of the damping of the radial pulsations associated with a certain radial
eigenmode in terms of the number of oscillation cycles:
\begin{equation}
N^{}_{\rm osc} = \frac{\tau^{(1,1)}_{lm}}{T^{}_{n}}\,,
\end{equation}
where $T^{}_{n} = \nu_{n}^{-1}\,,$ with $\nu^{}_{n}$ being the eigenfrequency of
the radial eigenmode.

Finally, it is important to make some comments on the difference between the
two scenarios we have studied: (i) scattering of gravitational waves
by a radially oscillating star, where we did not find any amplification or
resonant effect, and (ii) the coupling of differential rotation and radial
pulsations, where we have observed amplification when the frequencies associated
with these two different first order perturbations are close.
This difference is essentially due to the different properties of the source
terms in these two cases, which are determined by the first order radial and axial
non-radial perturbations.  In the first scenario, the source acts typically for a
relatively short period of time, which is given by the travel time of a
gravitational wave across the star.   After the gravitational waves have been scattered,
the first order signal still present inside the star decreases according
to the power law for gravitational wave tails.   As a consequence, there is
a ringing phase in the second order perturbations but it is well below the
amplitude of first order signal.
On the other hand, in the second scenario, the source terms act periodically on the star
forever, as this model does not contain the back-reaction.  As a result, the source
has more time to couple with the axial non-radial perturbations.

\section{Conclusions and Discussion}
\label{conclusions}
In this paper we have investigated non-linear effects in the dynamics
of compact stars due to the coupling of radial and axial non-radial oscillations,
and the potential impact that this mechanism may have in the emission
of gravitational radiation.   To that end, we have used and extended the
multi-parameter non-linear perturbative scheme introduced
in~\cite{Bruni:2002sm,Sopuerta:2003rg}, and further developed in~\cite{Passamonti:2004je}
for the case in which the coupling is between radial and polar non-radial
oscillations.

The non-linear perturbative equations and the gauge-invariant
character of the perturbative quantities describing the coupling
have been derived by using the 2-parameter perturbation theory in connection with
the GSGM formalism~\cite{Gerlach:1979rw,Gundlach:1999bt,Martin-Garcia:2000ze}.
As expected, in the stellar interior, the perturbative variables describing
the coupling satisfy inhomogeneous linear equations.  The form of these
equations is such that the associated homogeneous equations are constructed
from the same linear operators as the equations for  the first order
axial non-radial perturbations.   The source terms are quadratic in the
first-order radial and axial non-radial perturbative quantities.  In
the exterior we do not have source terms and hence, the dynamics is described
by Regge-Wheeler-type equations.  Interior and exterior communicate
through the junction conditions at the surface of the star.
We have discussed in detail the initial data setup used in the simulations
presented in this paper as well as the different boundary conditions.
The initial data for the radial oscillations corresponds to
specific eigenmodes.  In the case of the axial non-radial perturbations
we have used two types of initial data: (i) static profiles representing
differential rotation of the star from the relativistic $j$-constant rotation
law, and (ii) Gaussian packets containing a number of axial eigenmodes.
Regarding the boundary conditions, of particular importance are the conditions
at the stellar surface.   In order to avoid negative values of the mass energy
density near the surface due to the Eulerian description of the radial and
non-radial polar perturbations, we have adopted an approximation already used in
the literature~\cite{Sperhake:2001si,Sperhake:2001xi,2003PhRvD..68b4002H},
that consists in removing outer layers of the neutron star, which is equivalent,
in most of the cases we deal with,
to neglect less than one percent of the total gravitational mass of the star.
This approximation leads to a description of the second-order coupling
effects which is accurate to better than five per cent.

We have presented a computational framework for investigating, in the time
domain, the evolution of the axial non-linear perturbations describing the
coupling.  Our numerical codes are based on finite difference methods and
standard explicit evolution algorithms.  Their structure is hierarchical:
First, we solve the TOV equations; then, we evolve a time step the first-order
radial and non-radial perturbations, and with the result we update the
source terms for the non-linear perturbations; eventually, we evolve them 
in the same time step.

We have then studied two different physical scenarios: (i) The scattering of a
gravitational wave by a radially pulsating star.  (ii) A differentially rotating
and radially oscillating star.  The simulations for the first scenario have
shown that the correction onto the gravitational wave signal due to the
non-linear coupling are of the order of $2\,\%$ or less when the radial
oscillations are in the fundamental mode, and of the order of $0.1\,\%$
or less when we consider higher radial overtones.  Moreover, the properties
of the waveforms and spectra are essentially unchanged with respect to the
first-order ones.   The simulations for the second scenario have provided
more interesting results.  The waveforms we obtain have the following
pattern: at the early stages of the evolution they present an initial
$w$-mode burst related to the initial data choice. At later stages, 
the signal becomes periodic and is driven by the radial pulsations through 
the sources.  This is confirmed by an analysis of the spectra, which show 
that the gravitational wave signal contains the frequencies of the radial 
eigenmodes prescribed in the initial data.   More interestingly,  we have found 
a resonance effect that takes place as the frequencies of the radial eigenmodes 
considered get close to the frequency of the first $w$-mode. For the stellar 
model used in this paper, the amplitude of the gravitational wave signal associated 
with the fourth radial overtone (the one with the closest frequency to the first
$w$-mode) is about three orders of magnitude higher than that associated with
the radial fundamental mode.  We have also given a rough estimate of the damping
times of the radial pulsations due to the gravitational wave emission.
The values found depend strongly on the presence of resonances.  In the
case of differential rotation with a $10$ ms rotation period at the axis
and a differential rotation parameter $A=15$ km, the fundamental radial mode
gets damped after about ten billion oscillation periods, while the fourth
overtone after only ten.  This is not surprising, and shows that the coupling
near resonances is a very effective mechanism for extracting energy from the radial
oscillations.  In this sense, it is important to mention that the detection
of such a gravitational wave signature would provide important information
about the stellar physical properties, specially because it has information
on radial pulsations, which can be determined easily for a large class of equations
of state.

The work presented in this paper represents a first step in the study of the
non-linear coupling of oscillations of compact relativistic stars and its
impact for gravitational wave physics.  The next step in this study is the
description of the coupling between radial and polar non-radial oscillations.
The spectrum of the polar non-radial perturbations is richer than the axial case
due to the presence of the fluid modes, which may have frequencies lower than the
spacetime modes and a longer gravitational damping.   As a result, we may expect a
more effective coupling with the radial pulsation modes and more channels
for possible resonant amplifications.  Other extensions of this work include
the consideration of more realistic models of the stellar structure, by taking
into account the effects of rotation, composition gradients, magnetic fields,
dissipative effects, etc.  In particular, it would be interesting to investigate
a proto-neutron star model to compare the damping rates due to gravitational radiation
produced by the coupling of oscillations with the strong damping induced by the
presence of a high-entropy envelope surrounding the newly created neutron star.
By including rotation we may find new interesting non-linear effects due to the
different behaviour of radial and non-radial modes in a rotating configuration.
While the radial modes are only weakly affected so that their spectrum is
essentially the same as in a non-rotating configuration (when
scaled by the central density), the non-axisymmetric modes manifest a
splitting similar to the Zeeman effect in atomic spectra.
The rotation removes the mode degeneracy in the azimuthal harmonic number
of the non-rotating case.   The details of this frequency separation depend on
the stellar compactness and rotation rate.   It may then happen that for a given
stellar rotation rate and compactness the non-radial frequencies cross the sequence
of radial frequencies~\cite{Dimmelmeier:2005zk} so that the frequencies of these
two kinds of modes are comparable, and possible resonances or instabilities could
influence the spectrum and the gravitational wave signal.
Finally, as we have already mentioned above, it would be also interesting to
explore the impact of gravitational back-reaction on the dynamical coupling
mechanism (see, e.g.~\cite{Balbinski:85be,Yoshida:95ye}).

\[ \]
{\bf Acknowledgements:}
We thank Nils Andersson, Valeria Ferrari, Ian Hawke, Kostas Kokkotas,
Pablo Laguna, Giovanni Miniutti, Ulrich Sperhake and Nick Stergioulas 
for fruitful discussions.
AP work on this project is partly funded by a ``VESF'' grant. 
MB work was partly funded by a "Cervelli" fellowship from  MIUR (Italy).                   
CFS acknowledges the support of the Center for Gravitational Wave Physics
funded by the National Science Foundation under Cooperative Agreement
PHY-0114375, and support from NSF grant PHY-0244788 to Penn State University.

\appendix

\section{Gauge invariance \label{gaugeinvariance}}
In this section we show the gauge-invariant character of the $(1,1)$
perturbative quantities (\ref{ka_11_exp})-(\ref{L_11_exp})
(see for more details~\cite{Passamonti:2004je}).
Gauge transformations and gauge invariance in 2-parameter perturbation
theory have been studied in~\cite{Bruni:2002sm,Sopuerta:2003rg}.
The gauge transformation of a perturbation of order $(0,1)$ of
a generic tensorial quantity ${\cal T}$ is given by
\begin{equation}
\tilde{\cal T}^{(0,1)} = {\cal T}^{(0,1)} + \pounds^{}_{\xi^{}_{(0,1)}}{\cal T}^{(0,0)}\,,
\label{gafo}
\end{equation}
whereas a perturbation of order $(1,1)$ of ${\cal T}$ transforms
according to (see~\cite{Sopuerta:2003rg})
\begin{eqnarray}
\tilde{\cal T}^{(1,1)} & = &  {\cal T}^{(1,1)} +
\pounds^{}_{\xi^{}_{(0,1)}}{\cal T}^{(1,0)}
+\pounds^{}_{\xi^{}_{(1,0)}}{\cal T}^{(0,1)} {} \nn \\
{} & + &  \left(\pounds^{}_{\xi^{}_{(1,1)}} +
\left\{\pounds^{}_{\xi^{}_{(1,0)}}\,,\pounds^{}_{\xi^{}_{(0,1)}}
\right\} \right) {\cal T}^{(0,0)} \,, \label{gaugetrans11}
\end{eqnarray}
where $\{\,,\}$ stands for the anti-commutator $\{a,b\}= a\, b + b\,
a$.  In our work we fix the gauge for the radial $(1,0)$ perturbations.
Then, the transformation (\ref{gaugetrans11}) becomes
\begin{equation}
\tilde{\cal T}^{(1,1)} = {\cal T}^{(1,1)} +
\pounds^{}_{\xi^{}_{(0,1)}}\,{\cal T}^{(1,0)}
+\pounds^{}_{\xi^{}_{(1,1)}}\,{\cal T}^{(0,0)} \,. \label{ga_tran}
\end{equation}

To start with, let us expand in odd-parity vector harmonics
the generators $\xi^{}_{(0,1)}$ and $\xi^{}_{(1,1)}$ of the gauge
transformations:
\begin{equation}
\xi^{}_{(I,1)\alpha}=(0,0,r^2 V^{(I,1)}S^{}_a )\,,~~
I=0,1\,,\label{xi01}
\end{equation}
where $V^{(I,1)}$ are scalar functions on $M^2\,$.  Then,
the metric perturbations $h_A^{(I,1)}$ and $h^{(I,1)}\,$,
under the gauge transformation (\ref{gafo}), transform as follows:
\begin{eqnarray}
\tilde{h}_A^{(I,1)} & = & h_A^{(I,1)} + r^2 V^{(I,1)}_{\mid A}\,,
\label{ha_01_tr}\\
\tilde{h}^{(I,1)} & = & h^{(I,1)} + r^2 V^{(I,1)}\,.
\end{eqnarray}
and the $(1,1)$ energy-momentum tensor perturbations as
\begin{align}
& \delta \tilde{t}_A^{(1,1)} = \delta t_A^{(1,1)} + r^2\left[
\bar{p} V^{(1,1)}_{\mid A} + \bar{\rho}\bar{c}^2_s\omega^{(1,0)}
V_{\mid A}^{(0,1)}\right] \,, \\
& \delta \tilde{t}^{(1,1)} = \delta t^{(1,1)} + r^2\left[
\bar{p} V^{(1,1)} + \bar{\rho}\bar{c}^2_s\omega^{(1,0)} V^{(0,1)}\right]
\label{t_11_tr} \,.
\end{align}
Then, by using the transformation rules (\ref{ha_01_tr})-(\ref{t_11_tr})
it is straightforward to prove that the variables $k_A^{(1,1)}\,$,
$L_A^{(1,1)}\,$, and $L^{(1,1)}$ are gauge-invariant in the sense
explained above.  One can also apply the same procedure to study
the gauge invariance of the coupling perturbation $\Psi^{(1,1)}\,$.
By using the ansatz~(\ref{reorg}) in combination with its
expression~(\ref{Psibetadef}) at the $(0,1)$ order we get
\begin{eqnarray}
\Psi^{(1,1)} & = & \left[ r \left( k_{1\,,t}^{(1,1)} -
k_{0\,,r}^{(1,1)} \right) + 2 k_{0}^{(1,1)} \right] e^{-\left(\Phi +
\Lambda \right)} {} \nn \\
 {} & - & \eta^{(1,0)} \Psi^{(0,1)}  \,.
\end{eqnarray}
Since $\Psi^{(1,1)}$ depends linearly on the gauge-invariant
quantity $k_A^{(1,1)}$ and contains the product of the radial
perturbation $\eta^{(1,0)}$ with the first-order gauge-invariant
axial perturbation $\Psi^{(0,1)}\,$, we have that, for a fixed radial
gauge, the variable $\Psi^{(1,1)}$ is gauge-invariant, in the sense
that it is invariant under the transformations~(\ref{gafo})
and~(\ref{ga_tran}).

Finally, the gauge invariant character of the velocity perturbation
$\beta^{(1,1)}$ follows from the fact that $\bar{u}^{}_a=0$ and that
$\pounds^{}_{\xi^{}_{(0,1)}}u^{(1,0)}_{\alpha} =  0\,.$

\section{Source terms for the $(1,1)$ perturbation equations \label{sourceterms}}
In this Section we give the expressions of source terms $\Sigma^{}_{\Psi}$ and
$\Sigma^{}_{\beta}$ that appear in  the equations
(\ref{Psi11maseq})-(\ref{traseq11}) for the $(1,1)$ perturbations:
\begin{widetext}
\begin{eqnarray}
\Sigma^{}_{\Psi} & = &   2e^{-\Lambda}\left(rS^{(1,0)} - \eta^{(1,0)}\right)\Psi^{(0,1)}_{,rr}
+ \left\{ 4\pi (\bar\rho + \bar p) r \, \frac{1-\bar{c}_s^2}{\bar{c}_s^2} H^{(1,0)}
+ \left[ 4 \pi (\bar{p}-\bar\rho) r^2 -1 + \frac{4M}{r}\right]  S^{(1,0)} \right. \nn \\
& + & \left. 2 \left[ 4 \pi (\bar\rho -\bar p) r  - \frac{2M}{r^2} \right] \eta^{(1,0)} \right\} \Psi^{(0,1)}_{,r}
  - \left[ \eta^{(1,0)}_{,t} +  8 \pi (\bar \rho + \bar p ) r e^{\Lambda + \Phi }
  \gamma^{(1,0)} \right]  \Psi^{(0,1)}_{,t}  \nn \\
& + & \left\{ 4\pi (\bar\rho + \bar p ) \frac{1-\bar{c}_s^2}{\bar{c}_s^2} H^{(1,0)}
+ \left[ 4\pi (\bar p - \bar\rho) r - \frac{l(l+1)+3}{r} + \frac{12 M}{r^2}\right]S^{(1,0)} {} \right. \nn \\
& + & {} \left. 2 \left[ (\bar \rho - \bar p ) + \frac{l(l+1)}{r^2} - \frac{6M}{r^3} \right] \eta^{(1,0)} \right\}
\Psi^{(0,1)} - 8 \pi r e^{-\Lambda}  \left( 4 \eta^{(1,0)}  -  3 r S^{(1,0)} \right) \hat\beta^{(0,1)}_{,r}  \nn \\
& + & {} \left\{  24 \pi   \left(  4 \pi \bar p\,  r^3 + M \right) e^\Lambda  S^{(1,0)}  -  32 \pi
    \left( 4 \pi \bar p\, r^2 + \frac{M}{r} \right) e^{\Lambda} \eta^{(1,0)} +  16 \pi  r e^{-\Lambda}
    H^{(1,0)}_{,r}\right\}  \hat \beta^{(0,1)}\,,  \label{Spsi11} \\ \nn \\
\Sigma^{}_{\beta} & = &  -  e^{-\Lambda } \,\gamma^{(1,0)} \hat \beta^{(0,1)}_{,r}
   + e^{-\Lambda }\left\{ \left[ \left(4 \pi \rho \, r - \frac{M}{r^2} \right) \, e^{2
\Lambda}  - \frac{2}{r} \right] \gamma^{(1,0)}  - \gamma_{,r}^{(1,0)}\right\}
\hat\beta^{(0,1)} \,.  \label{Sbe11}
\end{eqnarray}
\end{widetext}

\section{Introducing the fluid tortoise coordinate \label{equationstortoise} }
Here we provide the expressions for the TOV equations and the equations for the
radial perturbations when we use the fluid tortoise coordinate
$x$ [see Eq.~(\ref{tort_fluid})].  The TOV equations (\ref{Phieq},\ref{meq}) become:
\begin{align}
& \bar{p}^{}_{,x} = -(\bar{\rho}+\bar{p})\Phi^{}_{,x} =
-\frac{\bar{c}^{}_s (\bar{\rho}+\bar{p})}{r-2M}\left(4\pi r^2\bar{p}+\frac{M}{r}\right)\,,
\label{tov1}
\\
& M^{}_{,x} = 4\pi r^2 \bar{\rho}\bar{c}^{}_s \,, \label{tov2}
\\
& r^{}_{,x} = \bar{c}^{}_s \,. \label{rtox}
\end{align}
The evolution equations (\ref{eq:H10_ev})-(\ref{chi_t}) for the perturbations
$H^{(1,0)}\,$, $\gamma^{(1,0)}\,$ and $S^{(1,0)}\,$, the equation for the metric
perturbation $\eta^{(1,0)}$~(\ref{eq:eta_cn}), and the Hamiltonian constraint
(\ref{eq:S10_cn}) take the following form:
\begin{widetext}
\begin{eqnarray}
H_{,t}^{(1,0)} & = & - \bar{c}^{}_s e^{\Phi-\Lambda}\gamma^{(1,0)}_{,x}
-\bar{c}_s^2 e^{\Phi + \Lambda}\left[\left(1-\frac{1}{\bar{c}_s^2}\right)\left(4\pi r\bar{p}
+\frac{m}{r^2}\right) + \frac{2}{r}e^{-2\Lambda}- 4\pi(\bar\rho+\bar{p})\right]\gamma^{(1,0)}\,,
\label{H10_ev_x}
\\
\gamma_{,t}^{(1,0)} & = & - \frac{e^{\Phi-\Lambda}}{\bar{c}^{}_s}H^{(1,0)}_{,x}
-4\pi r(\bar\rho + \bar{p})e^{\Phi + \Lambda}H^{(1,0)}
-\left(4\pi r^2\bar{p}+\frac{1}{2}\right)e^{\Phi+\Lambda}S^{(1,0)}\,,\label{gam_t_x}
\\
\eta^{(1,0)}_{,x} & = & 4\pi r\bar{c}^{}_s(\bar\rho+\bar{p})e^{2\Lambda}\left[rS^{(1,0)}
+\left(1+\frac{1}{\bar{c}^2_s}\right)H^{(1,0)}\right]\,, \label{eta_cn_x}
\\
S^{(1,0)}_{,x} & = & 2\bar{c}^{}_s e^{2\Lambda}\left(4\pi r\bar\rho -\frac{1}{r} +
\frac{m}{r^2}\right)S^{(1,0)} + 8\pi\frac{\bar\rho + \bar p}{\bar{c}^{}_s}e^{2\Lambda}H^{(1,0)}\,,
\label{S10_cn_x}
\\
S_{,t}^{(1,0)} & = & -8\pi(\bar\rho+\bar{p})e^{\Phi+\Lambda}\gamma^{(1,0)} \, .
\label{S_t_x} 
\end{eqnarray}
\end{widetext}
The equations of the eigenvalue problem for radial perturbations read
as follows:
\begin{eqnarray}
y_{,x}^{(1,0)} & = & \bar{c}^{}_s P^{-1} z^{(1,0)} \,, \label{yeq2}
\\
z_{,x}^{(1,0)} & = & -\bar{c}^{}_s\left(\omega^2 W + Q\right)y^{(1,0)}\,, \label{zeq2}
\end{eqnarray}
where $y^{(1,0)} = r^2 e^{-\Lambda} \gamma^{(1,0)} =
r^2 e^{-\Phi} \xi^r_{,t}$, and where the functions $W,P,Q$ are given by:
\begin{align}
& r^2 W = (\bar\rho+\bar{p})e^{3\Lambda+\Phi}\,,
\\
& r^2 P = (\bar\rho+\bar{p}) \bar{c}_s^2 e^{\Lambda+3\Phi}\,,
\\
& r^2 Q = (\bar\rho+\bar{p})\left[\frac{\Phi^{}_{,x}}{\bar{c}^{}_s}\left(
\frac{\Phi^{}_{,x}}{\bar{c}^{}_s}-\frac{4}{r}\right)
-8\pi\bar p e^{2\Lambda}\right]e^{\Lambda+3\Phi }\,.
\end{align}
The boundary conditions transform at the origin and on the surface
as follows:
\begin{equation}
y^{}_{0}  = \bar{c}^{}_s \frac{z^{}_{0}}{3 P} \,,  \qquad \left.
(\bar\rho + \bar p)\bar{c}^{}_s e^{-\Phi}
y^{(1,0)}_{,x} \right|^{}_{r=R_x}= 0\,. \label{bcx}
\end{equation}
%


\end{document}